\documentclass[pre,preprint,showpacs,preprintnumbers,amsmath,aps]{revtex4-1}
\usepackage[colorlinks=true,linkcolor=red,urlcolor=blue,citecolor=blue]{hyperref}
\usepackage{graphicx,hyperref}
\usepackage{xcolor}
\usepackage{bm}
\usepackage{subfigure}
\usepackage{amsthm}
\usepackage{axodraw2}
\usepackage{amsmath}
\usepackage{amssymb}

\newcommand\etab{\bar \eta}

\newcommand{\propfig}[2][0.55cm]{%
	\vcenter{\hbox{\includegraphics[height=#1]{#2}}}%
}

\newcommand{\diagfig}[2][1.75cm]{%
	\vcenter{\hbox{\includegraphics[height=#1]{#2}}}%
}

\def\nbC{{\mathchoice {\setbox0=\hbox{$\displaystyle\rm C$}%
			\hbox{\hbox to0pt{\kern0.4\wd0\vrule height0.9\ht0\hss}\box0}}
		{\setbox0=\hbox{$\textstyle\rm C$}\hbox{\hbox
				to0pt{\kern0.4\wd0\vrule height0.9\ht0\hss}\box0}}
		{\setbox0=\hbox{$\scriptstyle\rm C$}\hbox{\hbox
				to0pt{\kern0.4\wd0\vrule height0.9\ht0\hss}\box0}}
		{\setbox0=\hbox{$\scriptscriptstyle\rm C$}\hbox{\hbox
				to0pt{\kern0.4\wd0\vrule height0.9\ht0\hss}\box0}}}}
\def\nbQ{{\mathchoice {\setbox0=\hbox{$\displaystyle\rm
				Q$}\hbox{\raise
				0.15\ht0\hbox to0pt{\kern0.4\wd0\vrule height0.8\ht0\hss}\box0}}
		{\setbox0=\hbox{$\textstyle\rm Q$}\hbox{\raise
				0.15\ht0\hbox to0pt{\kern0.4\wd0\vrule height0.8\ht0\hss}\box0}}
		{\setbox0=\hbox{$\scriptstyle\rm Q$}\hbox{\raise
				0.15\ht0\hbox to0pt{\kern0.4\wd0\vrule height0.7\ht0\hss}\box0}}
		{\setbox0=\hbox{$\scriptscriptstyle\rm Q$}\hbox{\raise
				0.15\ht0\hbox to0pt{\kern0.4\wd0\vrule height0.7\ht0\hss}\box0}}}}
\def\nbT{{\mathchoice {\setbox0=\hbox{$\displaystyle\rm
				T$}\hbox{\hbox to0pt{\kern0.3\wd0\vrule height0.9\ht0\hss}\box0}}
		{\setbox0=\hbox{$\textstyle\rm T$}\hbox{\hbox
				to0pt{\kern0.3\wd0\vrule height0.9\ht0\hss}\box0}}
		{\setbox0=\hbox{$\scriptstyle\rm T$}\hbox{\hbox
				to0pt{\kern0.3\wd0\vrule height0.9\ht0\hss}\box0}}
		{\setbox0=\hbox{$\scriptscriptstyle\rm T$}\hbox{\hbox
				to0pt{\kern0.3\wd0\vrule height0.9\ht0\hss}\box0}}}}
\def\nbS{{\mathchoice
		{\setbox0=\hbox{$\displaystyle     \rm S$}\hbox{\raise0.5\ht0%
				\hbox to0pt{\kern0.35\wd0\vrule height0.45\ht0\hss}\hbox
				to0pt{\kern0.55\wd0\vrule height0.5\ht0\hss}\box0}}
		{\setbox0=\hbox{$\textstyle        \rm S$}\hbox{\raise0.5\ht0%
				\hbox to0pt{\kern0.35\wd0\vrule height0.45\ht0\hss}\hbox
				to0pt{\kern0.55\wd0\vrule height0.5\ht0\hss}\box0}}
		{\setbox0=\hbox{$\scriptstyle      \rm S$}\hbox{\raise0.5\ht0%
				\hboxto0pt{\kern0.35\wd0\vrule height0.45\ht0\hss}\raise0.05\ht0%
				\hbox to0pt{\kern0.5\wd0\vrule height0.45\ht0\hss}\box0}}
		{\setbox0=\hbox{$\scriptscriptstyle\rm S$}\hbox{\raise0.5\ht0%
				\hboxto0pt{\kern0.4\wd0\vrule height0.45\ht0\hss}\raise0.05\ht0%
				\hbox to0pt{\kern0.55\wd0\vrule height0.45\ht0\hss}\box0}}}}
\def\nbZ{{\mathchoice {\hbox{$\sf\textstyle Z\kern-0.4em Z$}}
		{\hbox{$\sf\textstyle Z\kern-0.4em Z$}}
		{\hbox{$\sf\scriptstyle Z\kern-0.3em Z$}}
		{\hbox{$\sf\scriptscriptstyle Z\kern-0.2em Z$}}}}

\makeatletter
\def\@fnsymbol#1{\ensuremath{\@alph{#1}}} 
\makeatother

\begin{document}
	
	\title{Activated dynamics in the quantum random field Ising model}
	
	\author{Ivan Balog} \email{balog@ifs.hr}
	\affiliation{Institute of Physics, P.O.Box 304, Bijeni\v{c}ka cesta 46, HR-10001 Zagreb, Croatia}
	
	\author{Lovro \v{S}aravanja} \email{lsaravanja@ifs.hr}
	\affiliation{Institute of Physics, P.O.Box 304, Bijeni\v{c}ka cesta 46, HR-10001 Zagreb, Croatia}
	
\author{Andrei A. Fedorenko} \email{andrey.fedorenko@ens-lyon.fr}
\affiliation{Univ Lyon, ENS de Lyon, CNRS, Laboratoire de Physique, F-69342 Lyon, France}
	
	\date{\today}

	\begin{abstract}
		
		We study the critical dynamics of the quantum random-field Ising model using the nonperturbative functional renormalization group (NP-FRG). The static critical behavior is found to be controlled by the zero-temperature fixed point of the classical random-field Ising model, where both thermal and quantum fluctuations are dangerously irrelevant. Considering a family of quantum dynamical universality classes defined by a bare dynamical kernel $F_\Lambda(\omega)\sim |\omega|^\sigma$, we show how this fluctuationless fixed point nevertheless controls the quantum dynamics by computing the full Matsubara-frequency dependence of the running dynamical kernel $F_k(\omega)$. This is essential at zero temperature: a naive treatment of the dynamical kernel flow leads to a divergence at a finite length scale, resulting in apparent localization. In contrast, keeping the full frequency dependence of the dynamical kernel and choosing a regulator adapted to its running scale yields a controlled flow. The resulting dynamics is of activated form, with a relaxation time given by $\ln \tau \sim \xi^\Psi$. The exponent $\Psi$ is determined by the static RFIM fixed-point exponents and by $\sigma$. 	At finite temperature, the flow crosses over to the classical thermally activated scaling of the random-field Ising model. These results provide a quantitative field-theoretic realization of the heuristic activation scenario proposed earlier for the quantum random-field model and establish a framework for analyzing the dynamics of other disordered quantum systems that may exhibit similar  tentative localization-like singularities.

	\end{abstract}
	
	\pacs{11.10.Hi, 64.60.Ht, 75.10.Nr, 75.40.Gb}

	\maketitle
	
	\section{Introduction}
	\label{sec:introduction}
	
\begin{figure}[]
	\centering
	\includegraphics[width=0.65\textwidth]{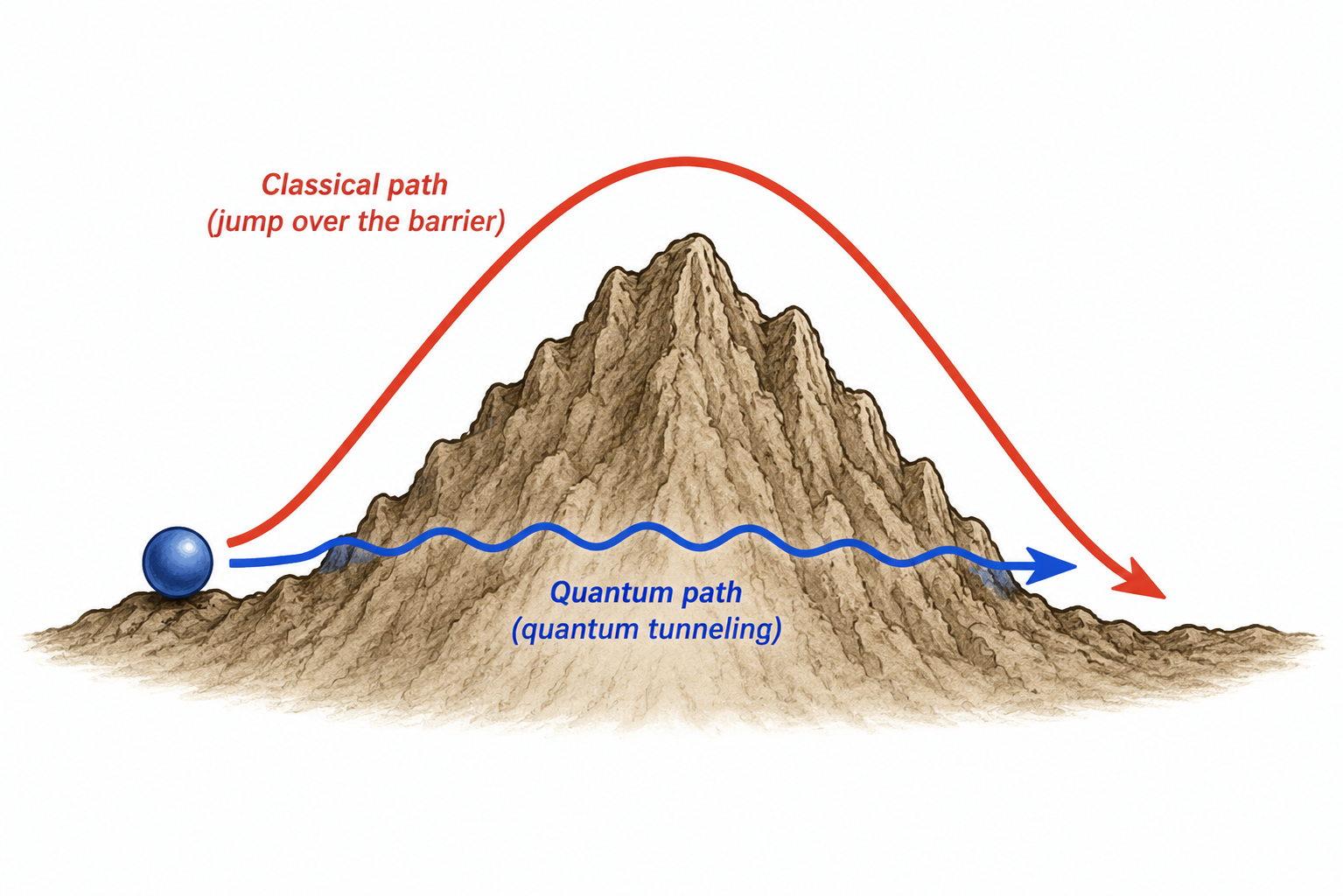}
	\caption{Schematic representation of a relaxation processes related with traversing a free energy barrier in the $T>0$ (classical) and the $T=0$ (quantum) cases.}
	\label{fig:barrier_cartoon}
\end{figure}

According to the seminal theory of dynamical critical phenomena, the relaxation time in the vicinity of a critical point scales  as $\tau \propto \xi^z$, where $\xi$ is the correlation length and $z$ is the dynamical critical exponent~\cite{halperin72,hohenberg77}.
%
This scaling picture can, however, break down in strongly disordered systems such as glasses~\cite{fisher88,braymoore87,lubchenko07,bouchaud98} and random-field magnets~\cite{villain84_b,fisher86,balog_activated}. In these systems, quenched disorder can give rise to rare-event effects and broad distributions of local observables, leading to Griffiths-type phenomena and \emph{activated} dynamical scaling~\cite{vojta06}.
In this regime, the system is no longer governed by fluctuations around a single free energy minimum. Instead, the free energy landscape becomes rough at large scales, with many metastable configurations separated by barriers that grow with the correlation length $\xi$ as a power law. Relaxation is therefore controlled by rare, collective rearrangements between these configurations, either via thermal activation at finite temperature $T$ or via quantum tunneling at $T=0$ (see Fig.~\ref{fig:barrier_cartoon}), leading to exponentially slow dynamics, with the relaxation time
\begin{equation}
	\label{eq:activated}
	\tau\propto e^{\textrm{const}\ \xi^{\Psi}}.
\end{equation}
While thermal activation over energy barriers can induce transitions between states whose energies significantly differ, quantum tunneling can effectively hybridize two configurations only when their energy mismatch is comparable to or smaller than the tunneling matrix element. Thus, the exponent $\Psi$ is determined primarily by the statistics of barrier heights in the classical case, whereas in the quantum limit it reflects the scaling of tunneling amplitudes and the associated distribution of low-energy excitation gaps. This implies that rare-event effects are more pronounced in the quantum regime, and consequently some models may exhibit activated dynamics only at zero temperature. A prominent example is the random transverse-field Ising model, where activated scaling has been established in one and two dimensions using strong-disorder renormalization-group methods~\cite{fisher92,fisher95,Motrunich2000}.

In the present work we study a class of systems where disorder is \textit{a priori} stronger. A prototypical example is the transverse-field Ising model with random longitudinal fields, or more simply the \emph{quantum random-field Ising model} (QRFIM), which provides a minimal setting in which random longitudinal fields compete directly with exchange interactions, transverse quantum fluctuations, and, at finite temperature, thermal fluctuations.

A typical realization of the classical RFIM is Fe$_x$Zn$_{1-x}$F$_2$, a randomly diluted uniaxial antiferromagnet in a uniform applied field in which strong experimental evidence was found of activated dynamics \cite{nash91,lederman92,belanger97}. A typical experimental realization of the QRFIM on the other hand is LiHo$_x$Y$_{1-x}$F$_4$, where dilution combined with a transverse magnetic field  induces \emph{effective longitudinal random fields} through dipolar interactions and symmetry-breaking effects, providing a tunable platform to access random-field physics in the presence of quantum fluctuations~\cite{tabei06,schechter08}. This and related materials exhibit coherent localized excitations and anomalously slow dynamics, highlighting the need for a controlled understanding of low-energy dynamical processes in disorder-dominated regimes~\cite{silevitch2007,silevitch19}.
It can also be realized in molecular magnets, where a transverse magnetic field couples to disorder in the local easy-axis orientations, effectively generating a tunable random-field Ising ferromagnet in a transverse field~\cite{wen10}. More recently, inhomogeneous altermagnets have been proposed as an additional platform, where a uniform magnetic field induces an effective random longitudinal field conjugate to the altermagnetic (Ising-like) order parameter via piezomagnetic coupling in the presence of random strain, thereby realizing a rare example of a field-tunable random-field Ising system~\cite{Chakrabot25}.
These systems provide strong motivation for a deeper understanding of the QRFIM, including the crossover from quantum to thermal criticality in its slow and glassy dynamics.

A reasonable heuristic picture of the critical scaling behavior of the QRFIM was developed by Senthil~\cite{senthil98}, who argued that in spatial dimensions $d>2$ the $T=0$ transition is controlled by a \emph{fluctuationless} (classical) fixed point, which also governs the finite-temperature transition of the classical RFIM. This picture calls into question numerous early studies~\cite{aharony82,boyanovsky83,MicnasChao1984} of the QRFIM which, by analogy with the RFIM, proposed a quantum version of dimensional reduction \cite{parisi79} (which was later proven to fail for the RFIM in three dimensions~\cite{bricmont87}). Within
this framework, the static critical exponents are expected to coincide with those of the classical RFIM, while the dynamics is predicted to be activated, governed by rare regions and tunneling between nearly degenerate minima. However, both the activated dynamics and the dominance of the classical fixed point have remained qualitative conjectures, supported only by heuristic and scaling arguments and lacking any explicit field-theoretic derivation or quantitative predictions.

The aforementioned classical RFIM problem has a long history (see Sec.~\ref{sec:historical_overview}), however only recently breakthroughs in the nonperturbative  functional renormalization group (NP-FRG) description \cite{tarjus04,tarjus04_a,tarjus04_b,tissier11,tissier12a,tissier12b} have put on firm analytical ground the notions of rugged free-energy landscapes~\cite{fisher86, fisher86b} and their implications for the analytical structure of the theory.

The present work is motivated by closing precisely the gap between the heuristic picture of~\cite{senthil98} and the modern nonperturbative approach \cite{tarjus04,tarjus04_a,tarjus04_b}. Building on the NP-FRG we provide an explicit calculation that (i) proves that both finite and zero temperature regimes of the QRFIM are controlled by the same "fluctuationless" fixed point of the RFIM and (ii) derives the activated dynamical scaling in both regimes and demonstrates a crossover between them at finite temperature.  This yields a sharper theoretical benchmark for interpreting experiments in tunable random-field quantum Ising platforms~\cite{tabei06,schechter08,wen10} and for connecting QRFIM criticality to the broader theory of rare-region--dominated disordered quantum phase transitions~\cite{vojta06}.
We also wish to stress a methodological progress within the NP-FRG framework (iii) brought by our work  since we for the first time to our knowledge establish a regulating scheme that enables a controlled evaluation of the Matsubara frequency dependence of the effective average action (i.e. the dynamical kernel) of the problem which is highly nontrivial since its renormalization is atypical as it retains feedback between low and high frequencies  along the entire RG flow.

The article is organized as follows. In Sec.~\ref{sec:formulation} we introduce the QRFIM, its continuum imaginary-time field theory, the multicopy disorder average, and the NP-FRG framework used throughout the work. We also specify the truncation and recall the classical RFIM fixed-point structure that controls the static sector. In Sec.~\ref{sec:flow_dyn_kernel_setup}, we formulate the classical fixed-point hypothesis and discuss its consequence: the cusp in the renormalized disorder cumulant is rounded at any finite RG scale within a boundary layer whose width shrinks along the flow. Using this boundary-layer structure, we  derive the leading flow of the dynamical kernel and introduce the frequency regulator needed to obtain a controlled flow. Sections~\ref{sec:finite_T_kernel} and~\ref{sec:zero_T_kernel} analyze the finite- and zero-temperature regimes, respectively, and extract the corresponding activated scaling laws. We also formulate the consistency criterion for the classical fixed point hypothesis. In Sec.~\ref{sec:numerics} we illustrate the analytical predictions with numerical solutions and discuss how insufficient frequency resolution can mimic localization-like behavior. The scaling results are collected in Sec.~\ref{sec:summary} where the classical fixed point hypothesis is confirmed, and Sec.~\ref{sec:conclusion} presents our conclusions and outlook. Technical details on subdominant quantum corrections and on the regulator diagnostics are deferred to the appendices.

		\section{Model and NP-FRG framework}
		\label{sec:formulation}
		
		\subsection{Model and the bare action}
		
\begin{figure}[]
			\centering
			\includegraphics[width=0.65\textwidth]{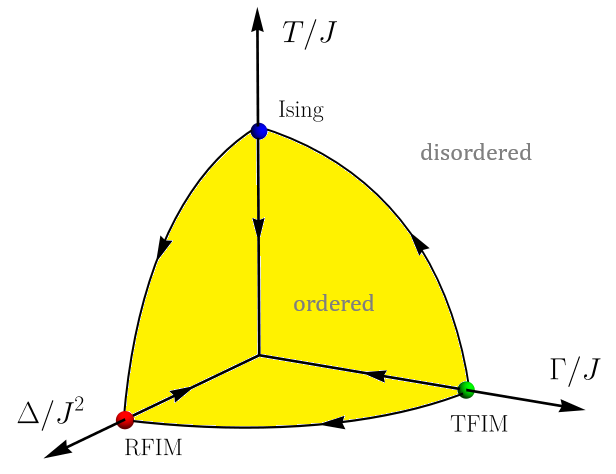}
			\caption{Schematic phase diagram and renormalization-group
				flows for QRFIM for $d>2$ in the dimensionless temperature $T$, transverse field $\Gamma$, and
				strength of random fields $\Delta$ space.  The fluctuationless  FP of the RFIM in $d$ dimensions attracts the entire critical surface, separating the ordered and disordered phases, whenever $\Delta>0$. The pure classical $d$ dimensional Ising model governs the transition at finite temperature and $\Delta=0$, while the the $d$-dimensional TFIM  fixed point is equivalent to the pure classical $d+1$-dimensional Ising model and governs the quantum transition of the pure model. }
			\label{fig:phase_diagram_TFIM}
		\end{figure}

		The model Hamiltonian that represents the QRFIM on a lattice \cite{tabei06,schechter08,silevitch19,wen10,Chakrabot25} is given by   
		\begin{equation}
			\hat H = - J \sum_{\langle i,j\rangle} \hat\sigma_i^z \hat\sigma_j^z
			- \Gamma \sum_{i} \hat\sigma_i^x
			- \sum_{i} h_i \hat\sigma_i^z,
			\label{eq:H_TFIM}
		\end{equation}
		where $\hat\sigma_i^{a}$ with $a = x,y,z$ are the Pauli matrices with site index $i$. The sum $\langle i,j \rangle$ runs over nearest-neighbor pairs of a $d$-dimensional lattice. The coupling $J>0$ denotes the ferromagnetic exchange interaction between neighboring spins, and $\Gamma$ is the transverse magnetic field, which induces quantum fluctuations by flipping spins along the $x$-direction. The longitudinal field $h_i$ acts along the $z$-direction and is taken to be a quenched, uncorrelated random variable with zero mean, $\overline{h_i}=0$, and variance $\overline{h_i h_j}=\Delta\delta_{ij}$. The expected phase diagram of the model~\eqref{eq:H_TFIM} is shown in Fig. \ref{fig:phase_diagram_TFIM}.

		The QRFIM \eqref{eq:H_TFIM} can be mapped, via the Suzuki-Trotter decomposition~\cite{suzuki76} followed by coarse-graining, onto
		a scalar field theory with quartic interaction and a quenched random field linearly coupled to the order parameter. Denoting by $\chi$ the fluctuating coarse-grained field, which depends on the $d$-dimensional coordinate $x$ and on imaginary time $\tau$, the corresponding Euclidean bare action takes the form
		\begin{eqnarray}
		\label{eq:bare_actionRFIM}
		S_{B}[\chi] &= & \int d\tau\int d^dx \Bigg\{ \frac{1}{2}(\partial^{\sigma/2}_\tau \chi(x,\tau))^2+\frac{1}{2}(\nabla \chi(x,\tau))^2 \nonumber\\
		&+& \frac{1}{2}r \chi(x,\tau)^2 +\frac{1}{4!}\lambda \chi(x,\tau)^4 - h(x)\chi(x,\tau)\Bigg\}
		\end{eqnarray}
	with $\sigma=2$. The fractional derivative allows for a general bare frequency dependence controlled by the exponent $\sigma$. This formulation describes a broader class of quantum dynamical universality classes~\cite{herz76, homenda24,homenda25}. In particular, the case $\sigma=1$ corresponds to overdamped quantum dynamics and arises, for example, in itinerant antiferromagnets, where the order-parameter fluctuations are damped through their coupling to particle-hole excitations of the underlying metallic bath~\cite{anfuso09}. 
		
		One can understand the phase diagram of the model through the mass parameter $r=r(\Gamma,J,\Delta)$ which measures the distance from criticality and is controlled by the microscopic couplings of the lattice model~\eqref{eq:H_TFIM}. At zero temperature, the model~\eqref{eq:bare_actionRFIM} can be viewed  as a classical
		RFIM in $d+1$ dimensions with random fields correlated along $\tau$ direction and uncorrelated in transverse directions. The quantum phase transition is reached when the renormalized mass vanishes, corresponding to $r=0$.

		At finite temperature, the imaginary-time direction is compact, $0 \le \tau \le \frac{1}{T}$, so that the field can be decomposed into Matsubara modes,
		\begin{equation}
			\chi(x,\tau) = \sum_{n} \chi_n(x) e^{i\omega_n \tau}, \qquad \omega_n = 2\pi n T.
		\end{equation}
		At sufficiently long wavelengths and for nonzero temperature, all modes with $n \neq 0$ acquire a finite mass and
		can therefore be integrated out. This leaves only the zero Matsubara mode $\chi_0(x)$, and the theory reduces to an effective classical action of the form
		\[
		S_{\mathrm{eff}} = \frac1{T} \int d^d x \, \left[ \frac{1}{2} (\nabla \chi_0)^2 + \frac{r_{\mathrm{eff}}}{2} \chi_0^2 + \frac{\lambda_{\mathrm{eff}}}{4!} \chi_0^4 - h(x) \chi_0
		\right],
		\]
		which is the standard classical RFIM theory in $d$-dimensions.
		After integrating out the quantum fluctuations, the effective couplings become temperature dependent,
		$r_{\mathrm{eff}} = r_{\mathrm{eff}}(T,\Gamma,J,\Delta)$  and the finite-temperature phase transition surface is determined by the condition $r_{\mathrm{eff}} = 0$. If $\Gamma$ is held fixed and temperature is varied, then close to this transition line one may write 		$r_{\mathrm{eff}} \simeq A \left[ T - T_c(\Gamma,J,\Delta) \right]$. 		Thus, the continuum description of the microscopic QRFIM crosses over from a $(d+1)$-dimensional classical RFIM with correlated disorder, where the mass parameter can be identified with $\Gamma - \Gamma_c$ at zero temperature, to a $d$-dimensional classical RFIM, where the mass parameter can be identified with $T - T_c$ at finite temperature.
		
		\subsection{Disorder averaging procedure}

		The key ingredient in Eq. \eqref{eq:bare_actionRFIM} is the random field $h(x)$ which is quenched in imaginary time. In the following we assume that it is drawn from a Gaussian distribution (with variance $\overline{h(x)h(y)}=\Delta_B\delta^d(x-y)$),
	\begin{equation}
		P[h]\sim \exp\left[-\frac{1}{2\Delta_B}\int d^d x h^2(x)\right].
	\end{equation}
	Following the multicopy construction of Tarjus and Tissier for the classical RFIM \cite{tarjus04,tarjus04_a,tarjus04_b}, we introduce independent copies before performing the disorder average. This gives the multicopy bare action
	\begin{eqnarray}
		\label{eq:bare_replicated_action}
		S^N_B[\{\chi_a\}]&=&\sum_a \int d\tau\int d^dx \Bigg\{ \frac{1}{2}(\partial^{\sigma/2}_\tau \chi_a(x,\tau))^2+\frac{1}{2}(\nabla \chi_a(x,\tau))^2 \nonumber\\ 
		&+& \frac{1}{2}r \chi_a(x,\tau)^2 +\frac{1}{4!}\lambda \chi_a(x,\tau)^4 \Bigg\}\nonumber\\
		&-& \frac{1}{2}\sum_{a,b}  \int d\tau\int d\tau'\int d^dx \Delta_B \chi_a(x,\tau)\chi_b(x,\tau') .
	\end{eqnarray}
We next introduce sources $J_a(x,\tau)$ for each copy and perform the Legendre transform,
	\begin{equation}
		\label{eq:partition_f}
	Z^N[\{J_a\}]=\int \Pi_a \mathcal{D}\chi_a e^{-S^N_B[\{\chi_a\}+ \sum_a\int d\tau\int d^dx J_a(x,\tau)\chi_a(x,\tau) ]},
	\end{equation}
which leads to the definition of the multicopy effective average action
\begin{equation}
	\label{eq:legendre_noreg}
\mathbf{\Gamma}[\{\phi_a\}]+\ln(Z^N[\{J_a\}])=\sum_a \int d\tau\int d^dx J_a(x,\tau)\phi_a(x,\tau) , 
\end{equation}
expressed in terms of the averaged fields 
\begin{equation}
	\label{eq:legendre_field}
	\phi_{\alpha}(x,\tau)=\langle \chi_{\alpha}(x,\tau)\rangle=\frac{\delta \ln(Z^N[\{J_a\}])}{\delta J_{\alpha}(x,\tau)}. 
\end{equation}
The multicopy effective action is naturally organized as an infinite expansion in disorder cumulants,
\begin{eqnarray}
	\label{eq:cumulant_expansion_Gamma}
\mathbf{\Gamma}[\{\phi_a\}] &=& \sum^{\infty}_{p=1}\frac{(-1)^{(p-1)}}{p!}\Gamma_{p,k}[\phi_{a_1},\cdots,\phi_{a_p}]\\
\Gamma_{1,k}[\phi_a]&=&\int d\tau\int d^d x \gamma_1[\phi_a(x,\tau)]\\
\Gamma_{2,k}[\phi_a,\phi_b]&=&\int d\tau \int d\tau' \int d^d x \gamma_2[\phi_a(x,\tau),\phi_b(x,\tau')]\\
\Gamma_{3,k}[\phi_a,\phi_b,\phi_c]&=&\int d\tau \int d\tau' \int d\tau'' \int d^d x \gamma_3[\phi_a(x,\tau),\phi_b(x,\tau'),\phi_c(x,\tau'')]\\
& & \cdots\nonumber.
\end{eqnarray}
The higher multicopy terms are the higher disorder cumulants. They are local in space but, because the disorder is quenched, the $n$-th cumulant generally depends on $n$ independent imaginary-time coordinates. This imaginary-time nonlocality is one of the central structural features of the the quantum problem. 

Note that the multicopy formalism that we use differs from the conventional replica trick \cite{MezardParisiVirasoro1987}, where all replicas are coupled to the same source and one therefore has access only to disorder cumulants evaluated for equal field arguments. Here, by considering multiple copies subjected to distinct sources that explicitly break the permutation symmetry between copies, we can determine the full functional dependence of the disorder cumulants.

\subsection{Effective average action and exact NP-FRG flow}
\label{sec:nprg_setup}

 We use the Wetterich effective average action formalism \cite{wetterich93,berges02,dupuis_review}. To integrate fluctuations progressively from short to long wavelengths, we add an infrared regulator to the bare action,
 \begin{equation}
 	\label{eq:regulator_generic}
 \Delta S_k[\chi_a,\chi_b]=\frac{1}{2}\int d\tau\int d\tau'\int d^dx\int d^dy \chi_a(x,\tau)R_{k;a,b}(x-y,\tau,\tau')\chi_b(y,\tau').
 \end{equation}

Apart from spatial translational invariance, the regulator is left generic at this stage. Its role is to suppress modes with $|q|\lesssim k$. Introducing $\Delta S_k$ in \eqref{eq:partition_f} and repeating the Legendre transformation gives the cutoff-dependent effective average action
\begin{equation}
		\label{eq:legendre_reg}
	\mathbf{\Gamma}_k[\{\phi_a\}]+\ln(Z^N_k[\{J_a\}])=\sum_a \int d\tau\int d^dx J_a(x,\tau)\phi_a(x,\tau) -\sum_{a,b} \Delta S_k[\phi_a,\phi_b],
\end{equation}
with $\Delta S_k[\phi_a,\phi_b]$ defined in \eqref{eq:regulator_generic}. The average fields are defined as in \eqref{eq:legendre_field}, but with the cutoff-dependent partition function. The scale evolution is governed by the exact Wetterich equation,
\begin{equation}
 \partial_k \mathbf{\Gamma}_k[\{\phi_a\}]=\frac{1}{2}\mathrm{Tr}\Bigg\{\partial_k R_{k}[\mathbf{\Gamma}^{(2)}_k+R_{k}]^{-1}\Bigg\},
\end{equation}
where the trace runs over space, imaginary time, and copy indices. We suppress the detailed arguments of the regulator for compactness, and $\mathbf{\Gamma}^{(2)}_k$ denotes the tensor of second functional derivatives with respect to the fields $\phi_a(x,\tau)$. The inverse operator $[\mathbf{\Gamma}^{(2)}_k+R_k]^{-1}$ is the full propagator and is understood as a tensorial expansion in copy space \cite{tarjus04_a},
\begin{equation}
\mathbf{G}_{k;x,y;\tau,\tau'}[\phi_a(x,\tau),\phi_b(y,\tau')]=\delta_{a,b}\hat{G}_{k;x,y;\tau,\tau'}[\phi_a(x,\tau)]+\tilde{G}_{k;x,y;\tau,\tau'}[\phi_a(x,\tau),\phi_b(x,\tau')]+\sum_c\cdots,
\end{equation}
where 
\begin{eqnarray}
	\hat{G}_{k;x,y;\tau,\tau'}[\phi_a(x,\tau)]&=& [\Gamma^{(2)}_{k,1}[\phi_a(x,\tau)]+R_{k,a,a}(x-y,\tau,\tau')]^{-1}|_{x,y;\tau,\tau'}, \\
	\tilde{G}_{k;x,y;\tau,\tau'}[\phi_a(x,\tau),\phi_b(x,\tau')]&=&\int_{x_1,x_2;\tau_1,\tau_2}	\hat{G}_{k;x,x_1;\tau,\tau_1}[\phi_a(x,\tau)](\Gamma^{(1,1)}_{k,2}[\phi_a(x_1,\tau_1),\phi_b(x_2,\tau_2)]\nonumber\\
	&-&R_{k,a,b}(x_1-x_2,\tau_1,\tau_2))|_{x_1,x_2;\tau_1,\tau_2}\hat{G}_{k;x_2,y;\tau_2,\tau'}[\phi_b(y,\tau')].
\end{eqnarray}

All cumulants remain local in space. The second and higher cumulants depend on independent imaginary times. Consequently, the one-copy propagator $\hat{G}$ carries the usual $q$ and $\omega$ dependence, while the two-copy propagator $\tilde{G}$ contains the zero-frequency filtering induced by the quenched disorder. For uniform average fields and after Fourier transforming in space and imaginary time, one obtains
\begin{align}
	\label{eq:ghatqw}
	\hat{G}_{k;q;\omega}[\phi_a]
	&=
	\left[
	\Gamma^{(2)}_{k,1}[\phi_a]+R_{k,a,a}
	\right]^{-1}\big|_{q;\omega}
	\equiv
	\propfig{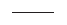},
	\\
	\label{eq:gtildeqw}
	\tilde{G}_{k;q;\omega}[\phi_a,\phi_b]
	&=
	\delta(\omega)
	\hat{G}_{k;q;\omega}[\phi_a]
	\left(
	\Gamma^{(1,1)}_{k,2}[\phi_a,\phi_b]
	-
	R_{k,a,b}
	\right)\big|_{q;\omega}
	\hat{G}_{k;q;\omega}[\phi_b]
	\equiv
	\propfig{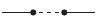}.
\end{align}


Both cumulants may contain spatial derivatives dependence, but the imaginary-time nonlocality of the disorder cumulant filters the two-copy propagator to the $\omega=0$ sector.

Using the diagrammatic notation in Eqs. \eqref{eq:ghatqw} and \eqref{eq:gtildeqw} and the shorthand notation $\tilde{\partial}_k$ for a derivative in $k$ only on the regulator functions, the flows of the first two cumulants can be written symbolically as
\begin{eqnarray}
	\label{eq:flow_Gamma1_general}
&&	\partial_k \Gamma_{1,k}[\phi_a]
	=
	\frac{\tilde{\partial}_k}{2}
	\mathrm{Tr}
	\left\{
	-\delta(\omega)
	\diagfig{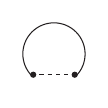}\right\}
	+\frac{1}{2}\mathrm{Tr}\left\{
	\diagfig{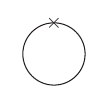}
	\right\},
\\
	&&	\partial_k \Gamma_{2,k}[\phi_a,\phi_b]
	=
	\frac{\tilde{\partial}_k}{2}
	\mathrm{Tr}
	\Bigg\{
	\delta(\omega)
	\Bigg(
	\diagfig{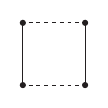}
	+
	2\diagfig{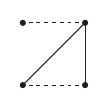}
	-
	2\diagfig{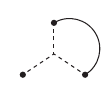}
	\Bigg)
	\nonumber\\
	&&
	\hspace{2.5cm}
	+
	2\diagfig{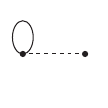}
	\Bigg\}. \label{eq:flow_Gamma2_general}
\end{eqnarray}
with the convention that copy index permutations are contained in the weight factors of the diagrams.  The "$\times$" in the second 1-copy diagram denotes that $\tilde{\partial}_k$ acted on the propagator giving an insertion of $\partial_k R_{k;a,a}$. The dashed line ending in dots signifies the second cumulant as in Eq. \eqref{eq:gtildeqw} and this symbolic notation generalizes straightforwardly to higher cumulants. 

In the diagrams that contain an inner loop that is closed by any disorder vertex - we call them the "disorder" diagrams - the Matsubara frequency trace is reduced to simply the $\omega=0$ term. On the other hand in the diagrams where the propagator loop is closed only by the first cumulant (hat) propagator, the Matsubara frequency trace remains - these diagrams we will call the "quantum" diagrams. It is important to realize that the disorder diagrams are fully classical diagrams since they do not contain the Matsubara frequency trace.

To make a connection with the work of Senthil \cite{senthil98}, if the physics is shown to be dominated by the disorder diagrams in the asymptotic limit at criticality, then the criticality will be dominated by the classical fixed point. Furthermore the flow equations derived solely from the disorder diagrams are identical to the flow equations for the classical equilibrium random field Ising model that are explained and studied in the works of Tarjus and Tissier \cite{tarjus04,tarjus04_a,tarjus04_b,tissier11,tissier12a,tissier12b}.  
 
We have introduced here the cutoff scale $k$, for convenience we shall use in the following interchangeably $k$ and $t=ln(k)$, where $t$ is the "renormalization time", not to be confused with real time or the imaginary time denoted consistently by $\tau$. Note that the renormalization time in our convention flows to negative values as $k$ flows to 0. In terms of RG time $k\partial_k=\partial_t$.  
 

\subsection{Truncation and fixed-point scaling}

Up to this point the construction has been exact. To obtain explicit flow equations, we truncate the cumulant hierarchy while retaining the distinction between disorder loops, which are filtered to $\omega=0$, and quantum loops, which contain a full Matsubara trace; see Eqs. \eqref{eq:flow_Gamma1_general} and \eqref{eq:flow_Gamma2_general}.

To obtain explicit flow equations we adopt a truncation of the cumulant hierarchy
and retain only the first two cumulants $\gamma_{1,k}$ and $\gamma_{2,k}$, while neglecting
$\gamma_{n\ge 3,k}$, 
\begin{eqnarray}
 &&\gamma_{1,k}[\phi_a(x,\tau)] = \frac{1}{2}\Big(F_k(\partial_\tau)  \phi_a(x,\tau)^2 +(\nabla \phi_a(x,\tau))^2Z_k[\phi_a(x,\tau)]\Big),\nonumber\\
	&& \hspace{23mm} +\ U_k[\phi_a(x,\tau)],\\
	&& \gamma_{2,k}[\phi_a(x,\tau),\phi_b(x,\tau')] = \phi_a(x,\tau)\phi_b(x,\tau')\Delta_k[\phi_a(x,\tau),\phi_b(x,\tau')],\\
	&&\gamma_{3,k}[\phi_a(x,\tau),\phi_b(x,\tau'),\phi_c(x,\tau'')] = 0.
\end{eqnarray}
Within this ansatz, the scale-dependent functions $U_k(\phi)$, $Z_k(\phi)$, and
$\Delta_k(\phi_a,\phi_b)$ are \emph{defined} as low-derivative projections of $\gamma_{1,k}$
and $\gamma_{2,k}$ (with $U_k$ the local potential, $Z_k$ the field renormalization, and
$\Delta_k$ the second disorder cumulant). 

The dynamical kernel $F_k(\partial_\tau)$ becomes a function $F_k(\omega)$ in Matsubara-frequency space. A central point of this work is that this function must be treated as a full frequency-dependent object, rather than expanded in small $\omega$. With external momentum $p$, the natural definition is
\begin{equation}
	\label{eq:dyn_kernel_definition}
	\Omega_k[\phi;\omega]\equiv\gamma^{(2)}_{1,k}[\phi;p=0,\omega]-\gamma^{(2)}_{1,k}[\phi;p=0,\omega=0],
\end{equation}
with
\begin{equation}
	\gamma^{(2)}_{1,k}[\phi;p,\omega]=\frac{\delta^2 \gamma_{1,k}[\phi]}{\delta\phi(p,\omega)\delta\phi(-p,-\omega)}.
\end{equation}
This quantity is in general field dependent. Guided by the classical RFIM dynamical analysis, where this field dependence is subleading \cite{balog_activated}, we work with $F_k(\omega)\equiv \Omega_k[\phi=0;p=0,\omega]$.

With such an ansatz and separation of the regulator into diagonal $R_{a,a}\equiv R_1$ and off-diagonal $R_{a,b}\equiv R_2$, while keeping the regulator dependence on both space and imaginary time, we obtain  the following expressions for the propagators
\begin{eqnarray}
	&& \hat{G}[\phi;q^2,\omega] = \frac{1}{R_1(q^2,\omega)+q^2Z(\phi)+F(\omega)+U''(\phi)},\\
	&& \tilde{G}[\phi_a,\phi_b;q^2,\omega] = \delta(\omega)\hat{G}[\phi_a,q^2,\omega](\Delta(\phi_a,\phi_b)+R_2(q^2,\omega))\hat{G}[\phi_b,q^2,\omega].
\end{eqnarray}
The corresponding flow equations are obtained by straightforward functional differentiation
of the exact diagrammatic flows \eqref{eq:flow_Gamma1_general} and \eqref{eq:flow_Gamma2_general}: (i) we evaluate the diagrams in uniform field configurations,
(ii) take the required field derivatives to project onto $U_k''(\phi)$ and $Z_k(\phi)$
(from the small external momentum expansion of the two-point vertex), and $\Delta_k$ (from mixed copy
derivatives), and (iii) expand the resulting vertices to the required order in external momentum.  The flow of the dynamical kernel (iv) is derived directly from the flow of $\gamma^{(2)}_{1,k}$ following the definition \eqref{eq:dyn_kernel_definition} (see the full expression in Appendix \ref{sec:appx_F_corrections}).

Once the truncation and projections are specified, the flow equations follow uniquely from the exact diagrams. For the static functions $U''$, $Z$, and $\Delta$, we write the dimensionful flows compactly as
\begin{eqnarray}
	\label{eq:flowUpp}
&&	k\partial_k U''_k(\phi)= \beta^{\star}_{U''} + \beta'_{U''}, \\
	\label{eq:flowZ}
&&	k\partial_k Z_k(\phi)= \beta^{\star}_{Z} + \beta'_{Z},\\
	\label{eq:flowDel}
&&	k\partial_k \Delta_k(\phi_a,\phi_b) = \beta^{\star}_{\Delta} + \beta'_{\Delta}.
\end{eqnarray}
We decompose each ``beta" function into disorder and quantum contributions, $\beta=\beta^\star+\beta'$. If the $\omega=0$ disorder sector dominates, retaining only the $\beta^\star$ terms reproduces the classical equilibrium RFIM flow equations (at the same approximation level). The classical parts are given in Refs. \cite{tissier12a,tissier12b}; the quantum corrections to $U''_k$ and $Z_k$ are listed in Appendix \ref{sec:appx_1copy_corrections}, while $\beta'_{\Delta}$ is discussed in Sec. \ref{sec:delta_rounding}.

We therefore begin from the hypothesis that the static critical point is the zero-temperature RFIM fixed point, and later check whether the quantum terms destabilize it; see Secs. \ref{sec:hypothesis_subdominance} and \ref{sec:criteria_consistency}. Under this hypothesis the relevant quantities scale as
\begin{eqnarray}
	\label{eq:sc1}
	{[} q {]} &=& k, \\
	{[}  \phi_{a,b}  {]} & =&k^{\frac{d-4+\bar{\eta}}{2}}, \\
		{[}  U  {]} & = &k^d,\\
	{[}  \gamma^{(2)}_{1,k} {]} &=& {[}  U''  {]} = {[}  q^2 Z  {]} = k^{2-\eta}, \\
	{[}  Z  {]} &= &k^{-\eta}, \\
	{[}  \Delta  {]} &= &k^{-(2\eta-\bar{\eta})},\\
	{[}  \hat{G}[\phi;q,\omega]  {]} &=& k^{-(2-\eta)}, \\
	\label{eq:scn}
   {[}  \tilde{G}[\phi_a,\phi_b;q,\omega]  {]} &=& k^{-(4+\bar{\eta})}.
\end{eqnarray}
In terms of dimensionless quantities defined by $U(\phi)=k^{d} u(\varphi)$, $Z(\phi)=Z_{0,k} z(\varphi)$ and $\Delta(\phi_a,\phi_b)=\Delta_{0,k} \delta(\varphi_a,\varphi_b)$, with $\phi_i=k^{\frac{d-4+\bar{\eta}}{2}}\varphi_i$, the flow equations read\\[-5mm]
\begin{eqnarray}
	\label{eq:u_dimless}
	&&\partial_k u''(\varphi) = -(2-\eta_k) u''(\varphi)+\frac{d-4+\bar{\eta}}{2}\varphi u''(\varphi) + \bar{\beta^{\star}_{u''}}+ \bar{\beta'_{u''}},\\
	\label{eq:z_dimless}
	&&\partial_k z(\varphi) = \eta_k z(\varphi)+\frac{d-4+\bar{\eta}}{2}\varphi z'(\varphi) + \bar{\beta^{\star}_z}+ \bar{\beta'_z},\\
	\label{eq:delt_dimless}
	&&\partial_k \delta(\varphi_a,\varphi_b) = (2\eta_k-\bar{\eta}_k)\delta(\varphi_a,\varphi_b)+\frac{d-4+\bar{\eta}}{2}(\varphi_a\frac{\delta}{\delta\varphi_a} +\varphi_b\frac{\delta}{\delta\varphi_b})\delta(\varphi_a,\varphi_b) + \bar{\beta^{\star}_\delta}+ \bar{\beta'_\delta},\ \ \ \ \ \ \
\end{eqnarray}
where the $\bar{\beta}$ designate the dimensionless flow functions. The flowing quantities $Z_{0,k}$ and $\Delta_{0,k}$ determine the (flowing) critical exponents $\eta_k$ and $\bar{\eta}_k$ by a set of 2 equations
\begin{eqnarray}
	&&\eta_k=-k\partial_k \ln(Z_{0,k}),\\
	&&\bar{\eta}_k=2\eta_k+k\partial_k \ln(\Delta_{0,k}).
\end{eqnarray}
The disorder cumulant scales as $\Delta\propto k^{-(2\eta-\bar{\eta})}$ and therefore diverges in the dimensionful quantities as $k\to0$. This follows from the Schwartz-Soffer inequality $2\eta-\bar{\eta}>0$ \cite{schwartz85}, which is respected by the NP-FRG description \cite{tarjus13a}.

A useful way to summarize the zero-temperature fixed point is that second-cumulant quantities dominate their first-cumulant counterparts:
\begin{equation}
	\left[\frac{\gamma^{(11)}_{2,k}}{\gamma^{(2)}_{1,k} }\right]=\left[\frac{\Delta}{U''}\right]=\left[\frac{\tilde{G}}{\hat{G}}\right]=\frac{1}{T_k}, 
\end{equation}
where $T_k\propto k^{\Theta_k}$ and $\Theta_k=2-\bar{\eta}_k+\eta_k$ is the running temperature exponent. Since $\Theta>0$, the temperature is dangerously irrelevant. More generally, the $n$-th disorder cumulant is enhanced by $T_k^{-(n-1)}$ relative to the first cumulant. 

\subsection{Dynamical kernel and relaxation scale}


We write the inverse propagator of the first cumulant as
\begin{equation}
	\gamma_{1,k}^{(2)}(q,\omega)
	=
	Z_k q^2 + U_k'' + F_k(\omega),
	\qquad
	F_k(0)=0 .
\end{equation}
The kernel $F_k(\omega)$ measures the cost of imaginary time-dependent fluctuations. It should not be directly identified with a real-time response function; rather, it provides the imaginary-time estimate of the tunneling or activation scale associated with rearranging a critical droplet. A simple energy-balance criterion is obtained by comparing the static cost at scale $k\sim L^{-1}$ with the frequency-dependent cost,
\begin{equation}
F_k(\omega_L)
\sim
Z_k k^2 + U_k'' .
\end{equation}

At criticality this defines a characteristic imaginary-frequency scale $\omega_L$, and hence
a heuristic relaxation time
\begin{equation}
\tau(L)\sim \omega_L^{-1}.
\end{equation}

If the growth of $F_k$ generated by the renormalization group is sufficiently rapid, this balance produces activated rather than power-law scaling.

There are two distinct cases. At $T=0$, imaginary frequency is a continuous variable, and the full function $F_k(\omega)$ is meaningful. If the low-frequency part of the bare kernel behaves as a power, for instance $F_\Lambda(\omega)\propto |\omega|^\sigma$, the running coefficient of this power determines how the characteristic frequency scale changes along the flow. In this case the dynamics is extracted from the renormalization of the full continuous imaginary-frequency kernel. 

At finite temperature, by contrast, imaginary time is compact and the frequencies are discrete,
\begin{equation}
\omega_n = 2\pi n T .
\end{equation}

The zero Matsubara mode is the static sector. The lowest nonzero Matsubara frequency, $\omega_1=2\pi T$, is therefore the first available probe of the dynamical kernel. In this case the analogue of the previous balance is not obtained by taking a continuous $\omega\to0$ limit, but by following the renormalization of $F_k(\omega_1)$. Thus the finite-temperature problem naturally turns the question of relaxation into the question of how the dynamical kernel evolves on the Matsubara grid.  

\subsection{Classical RFIM background}
\label{sec:historical_overview}

   Since a lot of our conclusions draw from the classical RFIM, we briefly recall the main historical developments related with it, which are most relevant for the present study.  
   
   The random-field Ising model (RFIM) emerged as a paradigmatic model for understanding the competition between ferromagnetic order and quenched local symmetry-breaking fields in a system of linear size $L$. The basic physical issue was clarified by the Imry-Ma argument, which balances the domain-wall cost proportional to $L^{d-1}$ against the random-field energy gain proportional to $L^{d/2}$, predicting impossibility to order at $d<2$ for Ising symmetry \cite{imry-ma75}, which was later rigorously shown also to be true even precisely at $d=2$ \cite{aizenman89,aizenman89_b}, confirming the lower critical dimension value of $d_{lc}=2$. Such heuristics was initially in tension with perturbative field-theoretic approaches, which led to dimensional reduction: the critical behavior of the RFIM in $d$ dimensions appeared identical to that of the pure Ising model in $d-2$ dimensions \cite{aharony76, grinstein76, parisi79}. The dimensional-reduction result was later understood as a consequence of an underlying supersymmetry in the formal continuum theory \cite{parisi79}, but it is now known to fail in low dimensions.
   
   A key rigorous milestone was the proof that the three-dimensional RFIM has long-range ordering at sufficiently weak disorder, validating the Imry-Ma lower-critical-dimension picture and ruling out the naive dimensional-reduction conclusion that there should not be a transition in $d=3$ \cite{imbrie84, bricmont87}. The physical mechanism behind the failure of dimensional reduction was connected to metastability, droplets, avalanches, and nonanalyticities in the renormalized disorder cumulants. Early phenomenological and scaling approaches emphasized the zero-temperature fixed point, the dangerously irrelevant temperature, modified hyperscaling, and activated dynamics \cite{villain84_b, fisher86}. Reviews by Nattermann and Belanger organized much of the theoretical and experimental status of the classical RFIM \cite{natterman98, birgenau98,belanger97}.  Several groups have convincingly demonstrated slow/glassy dynamics in Fe$_x$Zn$_{1-x}$F$_2$ \cite{nash91,lederman92,belanger97}, which is a typical experimental realization of the classical RFIM. 
   
   The perturbative renormalization group near the upper critical dimension $d_{uc}=6$ established the formal field-theoretic framework and the $\epsilon =6-d$ expansion, but also exposed the limitations of perturbation theory in describing realistic dimensions \cite{boyanovsky83}, especially regarding the exponentially slow, activated dynamics and the description of the zero-temperature fixed point. 
   
   Numerically, the RFIM became a benchmark for large-scale ground-state algorithms and finite-size scaling, with important work by Ogielski, Huse, Rieger, Young, Middleton, Fisher, Hartmann, and collaborators refining estimates of exponents and testing two-exponent scaling, universality, and zero-temperature criticality \cite{ogielski86, rieger93, hartmann02, middleton02}. The out-of-equilibrium RFIM under quasistatic driving also became a central model for hysteresis, Barkhausen noise, and crackling dynamics \cite{dahmen96, perkovic99, sethna01}.
   
   A major modern development was the nonperturbative functional renormalization group (NP-FRG) formulation of Tarjus and Tissier, which treats the full functional dependence of disorder cumulants and explains dimensional-reduction breakdown through the appearance of a cusp in the renormalized second cumulant of the random field \cite{tarjus04,tarjus04_a,tarjus04_b,tissier11,tissier12a,tissier12b}. Such behavior is easy to visualize by a redefinition of copy-diagonal and copy-off-diagonal dimensionless fields 
      \begin{eqnarray}
     && 	\phi =\frac{\phi_a+\phi_b}{2},\\
    &&	\delta \phi = \frac{\phi_b-\phi_a}{2}.
   \end{eqnarray}
The statement that the dimensionless fixed point second-cumulant function $\delta^{*}$ spontaneously develops a cusp at the fixed point states that for a small field $\delta \varphi$ it behaves like 
\begin{equation}
   	\delta^{*}(\varphi,\delta\varphi)\approx \delta^{*}_0(\varphi)+\delta^{*}_1(\varphi)|\delta\varphi|+\cdots.
 \end{equation} 
This result is crucial in the unified view of the zero-temperature fixed point, avalanches, droplets, supersymmetry breaking, and the special dimension $d_{DR}\approx 5.1$, above which dimensional reduction and supersymmetry are restored. 
   
A later NP-FRG extension to dynamics showed that the equilibrium RFIM has activated critical dynamics, with $ln\tau\propto \xi^{\psi}$, and found the barrier exponent $\psi=\Theta$, the temperature irrelevance exponent, below $d_{DR}$, while for $d>d_{DR}$, $\psi=\Theta-2\lambda$ - the barrier exponent is modified by the avalanche number exponent $\lambda$ \cite{tarjus13} in a way to make it vanish at $d=6$ \cite{balog_activated}. The same functional viewpoint also clarified the relation between equilibrium criticality and hysteresis criticality: below $d_{DR}$, they are governed by distinct fixed points, despite numerically close exponents, whereas above $d_{DR}$ they coincide \cite{balog_dynamics}.
   
The quantum RFIM, realized for instance as an Ising model in a transverse field and a random longitudinal field, inherits many of the classical properties but with an additional imaginary-time structure. Senthil argued that the $T=0$ transition of the transverse-field RFIM is controlled by the same fluctuationless fixed point as the classical finite-temperature RFIM, with quantum fluctuations dangerously irrelevant and activated rather than power-law dynamical scaling \cite{senthil98}. 
  
  More recent works on quantum creep in disordered elastic systems~\cite{Gorkhov02} and on $O(N)$ random-field systems amenable to perturbative functional RG treatment~\cite{andreanov14} emphasize that the analytic continuation and frequency structure are subtle: the relevant dynamical kernel may involve a broad or activated scale rather than a conventional finite dynamical exponent, and care is needed in distinguishing Matsubara-frequency scaling from real-time response. This makes the RFIM a continuing test case for disorder-dominated quantum criticality, functional renormalization group methods, and the emergence of activated dynamics beyond conventional quantum critical scaling.

		\section{Flow of the dynamical kernel}
		\label{sec:flow_dyn_kernel_setup}

		We now derive the effective flow equation for $F_k(\omega)$. The logic is sequential: first we formulate the classical-fixed-point hypothesis, then identify the boundary-layer rounding of the RFIM cusp, and finally use this rounded cusp to obtain the leading flow of the dynamical kernel.
		
		\subsection{Classical-fixed-point hypothesis}
		\label{sec:hypothesis_subdominance}
		
		The quantum solution can be built on the classical RFIM fixed point only if quantum corrections to the static functions vanish as $k\to0$. We use this as the starting hypothesis, derive its consequences, and formulate its consistency criteria in Sec.~\ref{sec:criteria_consistency}. The results will later show that those criteria are satisfied for all cases considered below $d_{u}=6$ (see Sec.~\ref{sec:summary}). 
		
		For the main argument we assume $d<d_{DR}$, where the static fixed point has a cusp in the dimensionless disorder cumulant. The modifications for $d>d_{DR}$ are summarized in Sec.~\ref{sec:summary}.
		
		The exact flows of the first two cumulants, together with the scaling in Eqs.~\eqref{eq:sc1}--\eqref{eq:scn}, imply three general facts:
	
	\begin{itemize}
			\item[1)] All disorder diagrams have the same nominal scaling.
			\item[2)] At the level of the integrand, one may estimate the propagator scaling from the $\omega=0$ limit. With this assignment, the quantum diagrams are suppressed by a factor $T_k$ before the Matsubara trace is evaluated.
			\item[3)] Quantum diagrams contain a full imaginary-frequency trace, whereas disorder diagrams are projected to the zero-frequency sector. 
		\end{itemize}
		
		This parallels the classical finite-temperature RFIM, where thermal diagrams are suppressed relative to disorder diagrams by the running temperature. The new issue in the quantum problem is the Matsubara trace and the Matsubara frequency dependence: it could in principle compensate the factor $T_k$. A central task is therefore to determine whether the trace changes the dominance of the disorder diagrams.
		
		The subdominance hypothesis is thus equivalent to requiring that, after the Matsubara trace, the quantum contributions remain smaller by some factor $\epsilon_k\to0$ as $k\to0$.  
		
		If this holds, the static fixed point is the classical RFIM fixed point. The dynamical sector remains nontrivial, because the flow of $F_k(\omega)$ does not reduce to the classical dynamical problem. The rest of the calculation is a self-consistency test of this classical-static but nonclassical-dynamical scenario.
		
		\subsection{Boundary-layer rounding of the disorder cusp}
			\label{sec:delta_rounding}
		
		To test the hypothesis we first examine the quantum contribution to the flow of the disorder cumulant, $\Delta$, defined in Eq. \eqref{eq:flowDel}:
	\begin{eqnarray}
		\label{eq:delta_q_flow}
		\beta'_{\Delta}&=\int_{q}\int_{\omega}\frac{\partial_t R_1(q^2,\omega)}{4}\Bigg(4\hat{G}[\phi-\delta \phi,q^2,\omega]^3\Big(q^2Z'(\phi-\delta \phi)+U^{(3)}(\phi-\delta \phi)\Big)\nonumber\\
		&\times\Big(\Delta^{(1,0)}(\phi,\delta \phi)-\Delta^{(0,1)}(\phi,\delta \phi)\Big)  \nonumber\\
		&+ 4\hat{G}[\phi+\delta \phi,q^2,\omega]^3\Big(q^2Z'(\phi+\delta \phi)+U^{(3)}(\phi+\delta \phi)\Big)\nonumber\\
		&\times\Big(\Delta^{(1,0)}(\phi,\delta \phi)+\Delta^{(0,1)}(\phi,\delta \phi)\Big)\nonumber\\
		&+2\Delta^{(1,1)}(\phi,\delta\phi)\Big(\hat{G}[\phi+\delta \phi;q^2,\omega]^2-\hat{G}[\phi-\delta \phi;q^2,\omega]^2\Big)\nonumber\\
		&+\Big(\Delta^{(2,0)}(\phi,\delta\phi)+\Delta^{(0,2)}(\phi,\delta\phi)\Big)\Big(\hat{G}[\phi+\delta \phi,q^2,\omega]^2+\hat{G}[\phi-\delta \phi,q^2,\omega]^2\Big)\Bigg).
	\end{eqnarray}
	The crucial term is proportional to $\Delta^{(0,2)}$. If it remains finite, it rounds the cusp of the classical RFIM fixed point. Under the subdominance hypothesis this rounding is confined to a narrow boundary layer around $\delta\phi=0$. We therefore isolate the two competing contributions in this region:
	\begin{equation}
		\label{eq:delta_rounding}
			k\partial_k\Delta(\phi,\delta \phi)\approx\underbrace{\beta_{\Delta,an}+\mathcal{A}_{1,k}(\phi) (\Delta^{(0,1)}(\phi,\delta \phi))^2}_{\beta^{\star}_{\Delta}}-\mathcal{A}_{2,k}(\phi)\Delta^{(0,2)}(\phi,\delta \phi) + O(\epsilon_k),
	\end{equation}
The dimensionful coefficients are
 \begin{eqnarray}
		\mathcal{A}_{1,k}(\phi)&= \frac{1}{2}\int_q \partial_t R_1(q^2,\omega=0)\hat{G}[\phi;q^2,\omega=0]^3,\\
		\mathcal{A}_{2,k}(\phi)&= \frac{1}{4}\int_q\int_{\omega} \partial_k R_1(q^2,\omega)\hat{G}[\phi;q^2,\omega]^2.
	\end{eqnarray}
Simple power counting shows that the $\mathcal{A}_1$ term has the expected classical RFIM scaling,
	\begin{equation}
		[\mathcal{A}_{1,k}(\Delta^{(0,1)}_{k})^2]=k^{-(2\eta-\etab)}.
	\end{equation}
The term $\mathcal{A}_2$  depends on the unknown dynamical kernel and is given by
	\begin{equation}
		\mathcal{A}_{2,k}(\phi)=\frac{1}{4}\int_q\int_{\omega}\frac{k\partial_k R_{1,k}(q^2,\omega)}{(R_{1,k}(q^2,\omega)+q^2Z_k(\phi)+U''_k(\phi)+F_k(\omega))^2}.
	\end{equation}
		This expression cannot be evaluated before the flow of $F$ is known.  
		
		If quantum corrections are subdominant, the cusp rounding vanishes as $k\to0$. In the boundary layer where the $\Delta^{(0,2)}$ term matters, we use the rounded form
		\begin{equation}
		\label{eq:delta_rounding_ansatz}
		\Delta_k(\phi,\delta \phi) \approx 
		\Delta_{0,k}^*(\phi) 
		+ \frac{\mathcal{A}_{2,k}^*(\phi)}{\mathcal{A}_{1,k}^*(\phi)}
		\left[
		1 - \sqrt{\,1 +
			\frac{\mathcal{A}_{1,k}^*(\phi)^2\, \Delta_{1,k}^*(\phi)^2}{\mathcal{A}_{2,k}^*(\phi)^2}\,
			\delta\phi^2 }
		\right]
		+ O\!\left( \epsilon_k\right).
	\end{equation}
	
   \begin{figure}[t!]
		\centering
		\includegraphics[width=0.65\textwidth]{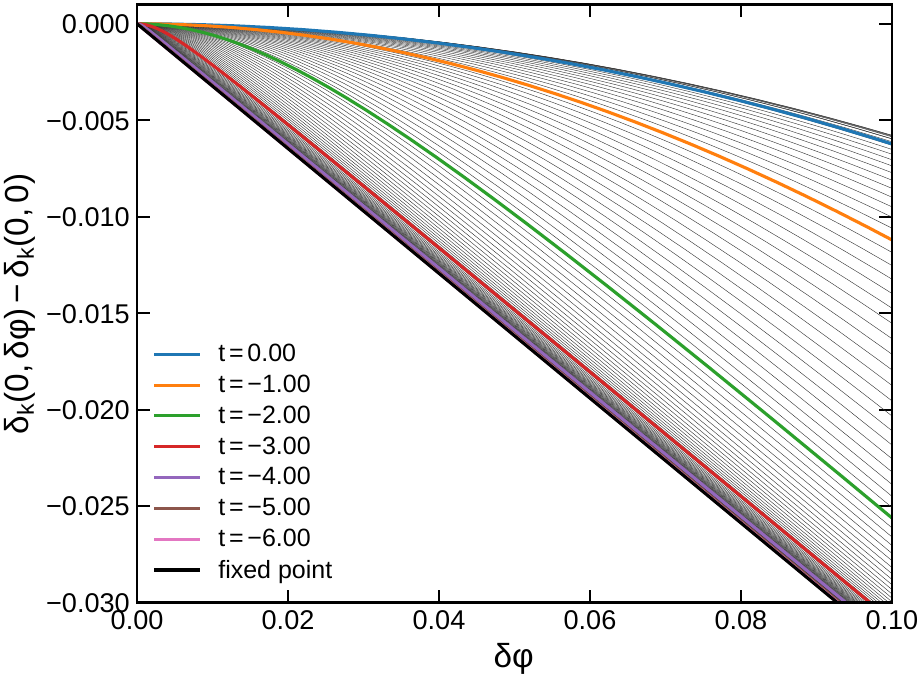}
		\caption{Thermally induced cusp rounding of the dimensionless disorder cumulant for the classical RFIM (in $d=4.4$). We expect the analogous scenario in the quantum case  where the width $\propto T_k$ in terms of dimensionless field $\delta\varphi$ is expected to be replaced by the width $\propto \epsilon_k$ (see Eq. \eqref{eq:subdominance_statement})				
		}
		\label{fig:BL_rounding}
	\end{figure}
	with
		\begin{eqnarray}
		\label{eq:Delta_0_scaling}
		\Delta^*_{0,k}(\phi)& =& k^{-(2\eta-\etab)}\delta^*_0(\varphi),\\
		\label{eq:Delta_1_scaling}
		\Delta_{1,k}^*(\phi)&=&k^{-(2\eta-\etab)-\frac{d-4+\etab}{2}}\delta^*_1(\varphi),
	\end{eqnarray}
	where the functions denoted by stars are fixed point functions (or constructed from the fixed point functions in the case of $\mathcal{A}$).  
	
	The subdominance condition can then be restated as
	\begin{equation} 
		\label{eq:subdominance_statement}
		\epsilon_k=\frac{\mathcal{A}_{2,k}^*(0)}{\mathcal{A}_{1,k}^*(0) \Delta_{0,k}^*(0)}\to 0  \quad\text{as }\quad k\to 0,
	\end{equation}
	where $\Delta_{0,k}^*(\phi)$ is the dimensionful fixed-point disorder cumulant at $\delta\phi=0$. When this condition holds, the boundary-layer width shrinks to zero in the dimensionless field $\delta\varphi=\delta\phi/k^{\frac{d-4+\bar{\eta}}{2}}$ as illustrated in Figure \ref{fig:BL_rounding} on the example of the thermally induced cusp rounding in the classical RFIM. When $\delta\varphi>>\epsilon_k$ the shape of the disorder cumulant function (see Eq. \eqref{eq:delta_rounding_ansatz}) is expected to converge to the fixed point function in the asymptotic limit, while for $\delta\varphi<<\epsilon_k$ it shows a rounded parabolic behavior. 
		
	The boundary layer is the bridge between the static RFIM cusp and the anomalous flow of the quantum dynamical kernel. We must therefore determine the dynamical kernel in order to close the argument.


		\subsection{Leading flow of the dynamical kernel}
	\label{sec:dyn_kernel}
	
	
	
	
		The full flow of $F$ is given in Appendix \ref{sec:appx_F_corrections} as well as the discussion of scaling of subdominant terms. Its leading disorder contribution contains $\Delta^{(0,2)}(\phi=0,\delta\phi=0)$. Equations \eqref{eq:subdominance_statement} and \eqref{eq:delta_rounding_ansatz} imply that this term is enhanced by the cusp boundary layer relative to the ordinary disorder terms which scale like $\gamma^{(2)}_{1,k}(q,0)\propto k^{2-\eta}$. We assume that the nonzero imaginary frequency sector scales as
	\begin{equation}
		\label{def:alpha_exp}
		\gamma^{(2)}_{1,k}(q,\omega\neq 0)\propto k^{\varkappa}, 
	\end{equation}
where the exponent $\varkappa$ is determined self-consistently below. The regime of interest has $k^\varkappa$ dominant over $k^{2-\eta}$ as $k\to0$.
		
	With most of the disorder and quantum diagrams as well subdominant, the flow of the dynamical kernel reduces to
	\begin{eqnarray}
		\label{eq:dyn_kernel0}
		\partial_t F_k(\omega) &\approx& -\frac{1}{4}\Delta^{(0,2)}_k(0,0)\int_q \Bigg(\frac{\partial_t R_{1,k}(q^2,\omega)}{(R_{1,k}(q^2,\omega)+q^2Z_k(0)+U_k''(0)+F_k(\omega))^2}\nonumber\\
		&-&\frac{\partial_t R_{1,k}(q^2,0)}{(R_{1,k}(q^2,0)+q^2Z_k(0)+U_k''(0))^2}\Bigg).
	\end{eqnarray}
	The subtraction guarantees $F(\omega=0)=0$.
		
		In the asymptotic region $k\ll1$, the static RFIM functions are replaced by their fixed-point forms. Inserting the boundary-layer ansatz \eqref{eq:delta_rounding_ansatz} then gives
	\begin{eqnarray}
		\label{eq:dyn_kernel1}
		\partial_t F_k(\omega) &\approx& \frac{1}{4}\frac{\mathcal{A}_{1,k}^*(0)\, \Delta_{1,k}^*(0)^2}{\mathcal{A}_{2,k}^*(0)}\int_q \Bigg(\frac{\partial_t R_{1,k}(q^2,\omega)}{(R_{1,k}(q^2,\omega)+q^2{Z^*}_k(0)+{U^*}_k''(0)+F_k(\omega))^2}\nonumber\\
		&-& \frac{\partial_t R_{1,k}(q^2,0)}{(R_{1,k}(q^2,0)+q^2{Z^*}_k(0)+{U^*}_k''(0))^2}\Bigg).
	\end{eqnarray}
	Equation~\eqref{eq:dyn_kernel1} is the central reduced flow: all nontrivial dynamics enters through the rounded cusp and the coefficient $\mathcal{A}_{2,k}^*$.
	
		\subsection{Choice of frequency regulator}
		
		
		We impose three sets of requirements on the regulator function.	The first requirement is standard \cite{litim01,canet03,balog19,depolsi22}: $R_k$ should suppress infrared modes while leaving ultraviolet modes unchanged. Thus $R_k(q)$ should be large enough to regularize the propagator for $q^2\ll k^2$, should rapidly vanish for $q^2\gg k^2$, should vanish as $k\to0$, and should respect the symmetries of the theory.
	
		The second requirement is specific to the RFIM: at zero frequency the regulator should respect the supersymmetric Ward identity relating the first and second cumulants. This implies 
	\begin{equation}
		\label{eq:susyreg}
		R_2(q,0)=-\partial_{q^2}R_1(q,0),
	\end{equation}
		so that the Ward identity is not explicitly broken \cite{tissier12a}. The two regulator components scale as
	\begin{eqnarray}
		{[}\! R_1 (q,0){]}\! &=& k^{2-\eta},\\
		{[}\! R_2 (q,0) {]}\! &=&  k^{-(2\eta-\etab)}.
	\end{eqnarray}
		When supersymmetry holds, $\eta=\etab$, and the two sides of Eq. \eqref{eq:susyreg} have compatible scaling. When supersymmetry is spontaneously broken, $\eta\neq\etab$; we impose Eq. \eqref{eq:susyreg} at the dimensionless level, while the dimensionful scaling follows the separate cumulant dimensions. 
		
		The third requirement concerns the frequency sector. We use the first-cumulant regulator
	\begin{equation}
		\label{eq:regulator_def}
		R_{1,k}(q,\omega)=c_0 k^2 Z_0 s_0(\frac{q^2}{k^2}+\frac{\omega^2}{\omega_{0}^2})+c_1 k^{\varkappa} s_1(\frac{q^2}{k^2},\frac{\omega^2}{\omega_{0}^2}).
	\end{equation} 
	The first term is the standard regulator, with the scaling of $\gamma^{(2)}_{1,k}(q,0)\sim Z_0k^2\propto k^{2-\eta}$. Similar regulators are common in nonperturbative approaches to frequency-dependent problems \cite{dupuis_review} . The second term is the new ingredient: it scales with the running dynamical kernel, $F\propto k^{\varkappa}$, and is designed to regularize only the dynamical part of the first cumulant. We refer to it below as the "special" regulator contribution. The constants $c_0$ and $c_1$ are regulator prefactors. Both regulator components depend on a characteristic frequency scale $\omega_0$, determined below (see Sec. \ref{sec:scaling_of_f}).

Some reasonable and previously discussed choices of the regulator functions for $s_0$ are 
	\begin{eqnarray}
		\label{eq:reg_exp}
		s_e(x) &=& e^{-x}, \\
		s_{\Theta}(x) &=& (1-x)\Theta(x-1).
	\end{eqnarray}
	The first choice will be referred to as the "exponential" regulator and the second choice as "$\Theta$" or "Litim" \cite{litim01} regulator (featuring the theta function which is 0  and 1 for arguments $>0$ and $<0$ respectively). 

	The function $s_1$ must vanish at $\omega=0$, matching the shape of the dynamical kernel and in addition not to modify the static sector. We choose
    \begin{equation}
    	\label{eq:s1regulator}
    	s_1(\frac{q^2}{k^2},\frac{\omega^2}{\omega_{0}^2})=\frac{\frac{\omega^2}{\omega_{0}^2}}{1+\frac{\omega^2}{\omega_{0}^2}}\cdot s_0(\frac{q^2}{k^2}), 
    \end{equation}
    with the same $s_0$ used in the static sector. 
    
    The absence of a large-frequency cutoff in $s_1$ is deliberate. It reflects the large-frequency behavior of the Eq. \eqref{eq:dyn_kernel1} discussed in Secs. \ref{sec:scaling_of_f} and \ref{sec:omega0}, and the large frequency regularization mechanism discussed in Sec. \ref{sec:scaling_kernel_T0} and Appendix \ref{sec:appx_regulator_insights}, rather than an arbitrary choice.
    
    \subsection{Scaling dimension of the dynamical kernel}
    \label{sec:scaling_of_f}
    
    
    We extract the scaling of the dynamical kernel from Eq. \eqref{eq:dyn_kernel1}. We use $y=q^2/k^2$, $\bar{\omega}^2=\omega^2/\omega_0^2$, and write $Z_0k^2\equiv k^{2-\eta}$. With the fixed-point scaling dimensions inserted, Eq. \eqref{eq:dyn_kernel1} becomes
    \begin{eqnarray}
    	\label{eq:dyn_kernel2}
    	\partial_t F(\omega)&= &\frac{v_d}{4}\frac{a^*_1{\delta^*_1}^2}{a_{2,k}}k^{-(2\eta-\bar{\eta})}\int y^{d/2-1}dy \Bigg\{\frac{e_0(y+\bar{\omega}^2)}{(g^{-1}_s+k^{-(2-\eta)}\Xi(\omega))^2}-e_0(y)g^2_s\nonumber\\ 
    	&+&\frac{k^{\varkappa-(2-\eta)}e_1(y,\bar{\omega}^2)}{(g^{-1}_s+k^{-(2-\eta)}\Xi(\omega))^2}\Bigg\}, 
    \end{eqnarray}
    with $v_d=\frac{1}{2^{d+1}\pi^{d/2}\Gamma(d/2)}$ and the following definitions 
    
    \begin{eqnarray}
    \label{eq:Xi}
    	 	\Xi(\omega)&=& k^{\varkappa} s_1(y,\bar{\omega}^2)+F(\omega),\\
    	\label{eq:e0_definition}
    	e_0(y+\bar{\omega}^2)&=&(2-\eta)s_0(y+\bar{\omega}^2)-2 (y+\partial_t \ln(\omega_0)\bar{\omega}^2) s'_0(y+\bar{\omega}^2),\\
    	\label{eq:e1_definition}
    	e_1(y,\bar{\omega}^2)&=&\varkappa s_1(y,\bar{\omega}^2)-2 y s^{(1,0)}_1(y,\bar{\omega}^2)-2\partial_t \ln(\omega_0)\bar{\omega}^2 s^{(0,1)}_1(y,\bar{\omega}^2),\\
    	g_s&=&\frac{1}{s_0(y+\bar{\omega}^2)+y z^*(0)+{u^*}''(0)},\\
    	a^*_1&=&\frac{v_d}{2}\int y^{d/2-1}dy e_0(y)g^3_s,
    \end{eqnarray}
    and 
    \begin{equation}
    \label{eq:a2k_def}
      a_{2,k}=\frac{v_d}{4}\int y^{d/2-1}dy\int d\omega \Bigg\{\frac{e_0(y+\bar{\omega}^2)}{(g^{-1}_s+k^{-(2-\eta)}\Xi(\omega))^2}+\frac{k^{\varkappa-(2-\eta)}e_1(y,\bar{\omega}^2)}{(g^{-1}_s+k^{-(2-\eta)}\Xi(\omega))^2}\Bigg\}.
    \end{equation}
    Note that the $e_0$ proportional contributions in Eqs. \eqref{eq:dyn_kernel2} and \eqref{eq:a2k_def} derive from the standard regulator term, i.e. the first term in  Eq. \eqref{eq:regulator_def} and the $e_1$ proportional contributions derive the second, i.e. the special regulator contribution.
    
    Assuming that $a_{2,k}$ scales like a power law, we define 
    \begin{equation}
    	[ a_{2,k} ] =k^{\rho_2}.
    \end{equation}
 

    Let us determine the scaling of the dynamical kernel by considering first the Eq. (\ref{eq:dyn_kernel2})  in the limit when $\omega>>1$.  Keeping in mind that the scaling dimension of the function $F\propto k^{\varkappa}$ is by our hypothesis expected to scale dominantly to the static functions scaling like $k^{2-\eta}$, we see that in particular $k^{\varkappa-(2-\eta)}$ will grow very rapidly as $k\to 0$. When $\omega>>1$,  Eq. (\ref{eq:dyn_kernel2}) reduces to
    \begin{equation}
    	\label{eq:dyn_kernel_largew}
    	\partial_t F(\omega)\approx -\frac{v_d}{4}\frac{a^*_1{\delta^*_1}^2}{a_{2,k}}k^{-(2\eta-\bar{\eta})}\int y^{d/2-1}dy e_0(y)g^2_s.
    \end{equation}
    The right hand side of the equation is negative, while the RG time flows to negative values by definition, implying that the flow will tend to increase the value of the function.  
    
    We wish to stress that this equation is quite atypical in the fact that the dynamical kernel flows for all RG times for all frequencies. If we remember the flows of dimensionless fixed point functions that describe the statics at some fixed momentum, what typically happens is that once the contributions from momenta above the fixed momentum are taken into the account and have renormalized it, the quantity stops flowing. The opposite is true here and it is so regardless of approximation. This is one of the reasons why we chose the special regulator contribution in the regulator definition (Eq. \eqref{eq:s1regulator}) not to cut in frequencies. This would seem unnatural given that the dynamical kernel always flows for all frequencies regardless of cutting or not the large frequencies with the regulator.  
    
    Since the only $k$ dependence in Eq. \eqref{eq:dyn_kernel_largew} comes from the obvious power and the scaling of  $a_{2,k}$, self-consistency requires a relation between the scaling of the dynamical kernel and that of $a_{2,k}$
    \begin{equation}
    \label{eq:alpha_beta}
        \varkappa =-(\rho_2+2\eta-\bar{\eta}).
    \end{equation}
   Defining $F(\omega)=k^{\varkappa}f(\omega)$,  we can write 
    \begin{equation}
    	\label{eq:fb_omega_large}
    	\partial_t f(\omega)\approx -\varkappa f(\omega)-c, 
    \end{equation}
   where
   \begin{equation}
    	\label{eq:c_asymptotic}
    	c=\frac{v_d}{4}\frac{a^*_1{\delta^*_1}^2}{\bar{a}_{2}}\int y^{d/2-1}dy e_0(y)g^2_s
    \end{equation}
    with $\bar{a}_{2,k}=k^{-\rho_2} a_{2,k}$. We can devise the asymptotic flow of the dynamical kernel for large frequencies since the solution of the above equation is 
    \begin{equation}
    	f_k(\omega)=-\frac{c}{\varkappa}+\left(\frac{k}{\Omega}\right)^{-\varkappa}\left(f_0(\omega)+\frac{c}{\varkappa}\right)
    \end{equation} 
 with $\varkappa$ strictly negative and $\Omega$ the UV cutoff scale. The interpretation of this solution is that for large frequencies, the dimensionless dynamical kernel approaches a positive constant determined by $c$. Furthermore, the information about the initial condition of the dynamical kernel - $f_0(\omega)$ decays with a power law $k^{|\varkappa|}$ compared with the constant contribution. 

	 \subsection{Characteristic frequency \texorpdfstring{$\omega_0$}{omega0}}
\label{sec:omega0}

	The characteristic frequency is fixed by the decay of the UV initial condition in the large-frequency regime. If
\begin{equation}
	f_0(\omega) \propto \omega^\sigma
\end{equation}
with $\sigma$ typically equal to 1 or 2, the corresponding frequency scale is
\begin{equation}
	\omega_0\propto k^{\frac{\varkappa}{\sigma}}.
\end{equation}
This is the only power-law frequency scale produced by the large-frequency solution. Since $\varkappa<0$ in the regimes of interest, $\omega_0$ grows as $k\to0$. This is opposite to the usual momentum-sector intuition: modes at large dimensionless frequency continue to feed back into the flow rather than simply decoupling. Similar UV sensitivity was emphasized in Ref. \cite{Gorkhov02}. In a real system, this growth stops once $\omega_0$ reaches the microscopic UV scale; the universal regime discussed below is the regime before that cutoff is reached.  
	 
	 The growth of $\omega_0$ is also a reason why the $s_1$ regulator in Eq. \eqref{eq:s1regulator} was not chosen to cut off large dimensionless frequencies.

	     The remaining question is the scaling of $a_{2,k}$. This scaling depends on whether $T$ is finite or zero, and therefore gives different values of $\rho_2$.

	    \section{Finite-temperature dynamical kernel}
	    \label{sec:finite_T_kernel}
		
		\subsection{Scaling at finite temperature}
		At finite $T$, the Matsubara frequency integral becomes a sum, $\int d\omega \to T \sum_{\omega_n=2\pi T n}$. Accordingly, the dynamical kernel is defined on the discrete Matsubara grid. The coefficient $a_{2,k}$ becomes
	\begin{equation}
		\label{eq:a2k_Tfin}
		a^T_{2,k}=\frac{v_d T}{4}\int y^{d/2-1}dy\sum_{\omega_n=2\pi T n} \Bigg\{\frac{e_0(y+\bar{\omega}_n^2)}{(g^{-1}_s+k^{-(2-\eta)}\Xi(\bar{\omega}_n))^2}+\frac{k^{\varkappa-(2-\eta)}e_1(y+\bar{\omega}_n^2)}{(g^{-1}_s+k^{-(2-\eta)}\Xi(\bar{\omega}_n))^2}\Bigg\},
	\end{equation}
	where $\omega_n=2\pi Tn$ and $T$ is the true temperature (independent of $k$). Asymptotically, the sum is dominated by the zero Matsubara mode. Since only the standard regulator contributes at $\omega=0$, Eq. \eqref{eq:a2k_Tfin} reduces to
	\begin{equation}
		a^T_{2}\approx\frac{v_d T}{4}\int y^{d/2-1}dy e_0(y)g^2_s.
	\end{equation}
Thus, $a^T_{2}$ has scaling exponent $\rho_2^T=0$: it is asymptotically independent of $k$.  Using Eq.~\eqref{eq:alpha_beta}, the scaling dimension of the dynamical kernel is
\begin{equation}
	\label{eq:alpha_Tfinite}
\varkappa=-(2\eta-\bar{\eta}).
\end{equation}
This is the scaling of the second, i.e. disorder cumulant $\Delta$, even though $F_k$ belongs to the first cumulant. The nonzero-frequency first-cumulant sector is therefore governed by disorder scaling. 

\subsection{Small-frequency activated regime at finite temperature}

We now consider frequencies for which the dimensionful kernel satisfies $F(\omega)\ll1$. In this regime Eq. \eqref{eq:dyn_kernel2} reduces asymptotically to
\begin{eqnarray}
	\label{eq:dyn_kernel3T}
	\partial_t F(\omega)&\approx&-F(\omega)\frac{v_d}{2}\frac{a^*_1{\delta^*_1}^2}{a^T_{2}}k^{-(2-\bar{\eta}+\eta)} \int y^{d/2-1}dy e_0(y) g^3_s. 
\end{eqnarray}
The only explicit cutoff dependence is $k^{-(2-\bar{\eta}+\eta)}=k^{-\Theta}$, where $\Theta=2-\bar{\eta}+\eta$ is the temperature irrelevance exponent. Hence, while $F(\omega)\ll1$, the solution has activated form,
\begin{equation}
F_k(\omega)\propto e^{k^{-\Theta}}.
\end{equation}
This result is completely analogous to the expected behavior of the classical RFIM that was previously hypothesized by Fisher \cite{fisher86} and obtained via NP-FRG for the classical RFIM with Langevin dynamics \cite{balog_activated}. At finite temperature, the zero Matsubara mode restores the classical activated result with barrier exponent $\Psi=\Theta$.




		\section{Zero-temperature dynamical kernel}
			\label{sec:zero_T_kernel}
		
		\subsection{Scaling at zero temperature}
		\label{sec:scaling_kernel_T0}
		
		At $T=0$ the frequency integral remains continuous, and the scaling of $a_{2,k}$ differs from the finite-temperature case. We evaluate Eq. \eqref{eq:a2k_def} using the characteristic scale $\omega_0\propto k^{\varkappa/\sigma}$.  
		
		We first consider the second term, which contains the special frequency-dependent regulator in Eq. \eqref{eq:regulator_def}. Using the large-frequency form $F(\bar{\omega})\approx k^{\varkappa}(f_0+f_1 \bar{\omega}^{\sigma})$, and the fact that the dynamical contribution dominates the propagator at finite frequency, one obtains
	\begin{equation}
	\label{eq:regulating_term}
	\int y^{d/2-1}dy\int d\omega \frac{k^{\varkappa-(2-\eta)}e_1(y,\bar{\omega}^2)}{(g^{-1}_s+k^{-(2-\eta)}\Xi(\bar{\omega}))^2}\approx k^{\frac{\varkappa}{\sigma}-(\varkappa-(2-\eta))} \int y^{d/2-1}dy\int d\bar{\omega} \frac{e_1(y,\bar{\omega}^2)}{(f_0+c_1 +f_1 \bar{\omega}^\sigma )^2}.
	\end{equation}
	Using Eq. \eqref{eq:alpha_beta}, this term has scaling dimension $\frac{\sigma-1}{\sigma}(\rho_2+2\eta-\bar{\eta})+(2-\eta)$.
		
	The first term, which comes from the standard regulator, gives
	\begin{equation}
		\label{eq:nonregulating_term}
		\int y^{d/2-1}dy\int d\omega \frac{e_0(y+\bar{\omega}^2)}{(g^{-1}_s+k^{-(2-\eta)}\Xi(\bar{\omega}))^2}\approx k^{\frac{\varkappa}{\sigma}-2(\varkappa-(2-\eta))} \int y^{d/2-1}dy\int d\bar{\omega} \frac{e_0(y+\bar{\omega}^2)}{(f_0+c_1 +f_1 \bar{\omega}^\sigma )^2},
	\end{equation}
	  and therefore has scaling dimension $\frac{2\sigma-1}{\sigma}(\rho_2+2\eta-\bar{\eta})+2(2-\eta)$.
		 
		 Self-consistency requires the scaling exponent of $a_{2,k}$ to be $\rho_2$. The consistent asymptotic solution is obtained by matching $\rho_2$ to the scaling dimension of the second, frequency-regulating term, giving
	 \begin{equation}
	 \label{eq:beta_consistency}
	 \rho_2=(\sigma-1)\Theta+(2-\eta), 
	 \end{equation} 
	while the first term is subleading. A consistent solution could not have been obtained if $\rho_2$ was matched to the scaling of the first, i.e. standard regulator term since this would lead to the divergent behavior discussed in Sec. \ref{sec:cautionary_tales}. The special $s_1$ regulator is therefore essential at $T=0$: it produces a consistent scaling of $a_{2,k}$ and prevents the divergent behavior. 
   
   
   This mechanism resembles the regularization found by Gorokhov et al. \cite{Gorkhov02}: although the formalism is different, $a_{2,k}$ is again controlled by the large-frequency part of the integral \eqref{eq:regulating_term}. 
   
   Combining Eq. \eqref{eq:beta_consistency} with Eq. \eqref{eq:alpha_beta} gives the zero-temperature scaling of the dynamical kernel,
    \begin{equation}
   	\label{eq:alpha_T0}
   	\varkappa=-\sigma\Theta.
   \end{equation}
   
		
		\subsection{Small-frequency activated regime at zero temperature}
		\label{sec:dyn_kernel_T0}
		
		For fixed dimensionful frequencies with $F(\omega)\ll1$, Eq. \eqref{eq:dyn_kernel2} gives
	\begin{eqnarray}
		\label{eq:dyn_kernel30}
		\partial_t F(\omega)&\approx&-F(\omega)\frac{v_d}{2}\frac{a^*_1{\delta^*_1}^2}{\bar{a}^0_{2}}k^{-(\rho_2+\Theta)} \int y^{d/2-1}dy e_0(y) g^3_s 
	\end{eqnarray}
	with $a^0_{2,k}=k^{\rho_2}\bar{a}^0_{2}$. Therefore, while $F(\omega)\ll1$,
	\begin{equation}
		F_k(\omega)\propto e^{k^{-\Psi}}.
	\end{equation}
	At zero temperature, the continuous frequency sector gives a genuinely quantum activation dynamics with the barrier exponent $\Psi=\sigma \Theta+2-\eta$. 
	
		\subsection{Consistency criterion}
		\label{sec:criteria_consistency}
		
		If $a_{2,k}$ follows a power law, the subdominance condition \eqref{eq:subdominance_statement} becomes $\epsilon_k\propto k^{\Theta+\rho_2}\to0$ as $k\to0$. Thus, the construction is consistent provided that
	\begin{equation}
		\label{eq:validity_criterion}
		\rho_2>-\Theta.
	\end{equation}
	Equivalently, using the small-frequency result of Sec.~\ref{sec:dyn_kernel_T0}, this criterion implies
	\begin{equation}
		\label{eq:validity_criterion_+}
		\Psi>0.
	\end{equation}
	The same condition also guarantees the validity of the simplified flow equation. Indeed, Eq. \eqref{eq:alpha_beta} shows that even if $\varkappa$ becomes positive, the dynamical term still dominates the static one provided $\varkappa<2-\eta$. This bound is again equivalent to $\rho_2>-\Theta$ by virtue of the Eq.~\eqref{eq:alpha_beta}, and therefore to Eqs. \eqref{eq:validity_criterion} and \eqref{eq:validity_criterion_+}.

	Thus, Eqs. \eqref{eq:validity_criterion} and \eqref{eq:validity_criterion_+} are the asymptotic validity criteria for the calculation. They ensure classical fixed-point dominance, consistency of the cusp-rounding boundary layer, and validity of the leading dynamical-kernel flow in Sec. \ref{sec:dyn_kernel}. The whole construction is self-consistent precisely when the resulting activation exponent is positive.  
	
\section{Numerical demonstrations}
\label{sec:numerics}

We now illustrate the analytic scenario numerically. The goal is not to study pre-asymptotic behavior, but to verify the asymptotic scaling and the regularization mechanism. We thus solve the asymptotic flow equation \eqref{eq:dyn_kernel2}, using the RFIM fixed-point solution for the static functions. We use the exponential regulator (see Eq. \eqref{eq:reg_exp}) for the results obtained below.

Because the characteristic frequency grows as a power law, the large-frequency sector must be included accurately. We therefore map the infinite frequency domain to a finite one via
\begin{equation}
 \bar{\omega}=\frac{\upsilon}{1-\upsilon},
\end{equation}
so that $\upsilon\in[0,1]$ covers $\bar{\omega}\in[0,\infty)$. We parametrize $F$ on a uniform $\upsilon$ grid and evaluate the frequency integral/sum effectively on the full domain. To improve numerical stability we evolve the logarithm of the dimensionless kernel. If $\partial_t F(\omega)=k^{\varkappa}\bar{\beta}_F$ and $\omega_0\propto k^{\varkappa/\sigma}$, then
\begin{equation}
\label{eq:flow_lnf}
\partial_t \ln(f(\bar{\omega}))= -\varkappa +\frac{\varkappa}{\sigma}\bar{\omega} \partial_{\bar{\omega}} \ln(f(\bar{\omega})) +\frac{\bar{\beta}_F}{f(\bar{\omega})}.
\end{equation}
We integrate Eq.~\eqref{eq:flow_lnf} with an implicit Euler scheme. As a first check, we fix $\varkappa$ to the analytic values in Eqs. \eqref{eq:alpha_Tfinite} and \eqref{eq:alpha_T0}. The flow then recovers the expected asymptotic behavior: Eq. \eqref{eq:beta_consistency} at $T=0$ and $\rho_2=0$ at finite $T$. This verifies the regularization mechanism, but it uses the analytic conclusions as input.

\begin{figure}[h!]
	\centering
	\includegraphics[width=0.496\textwidth]{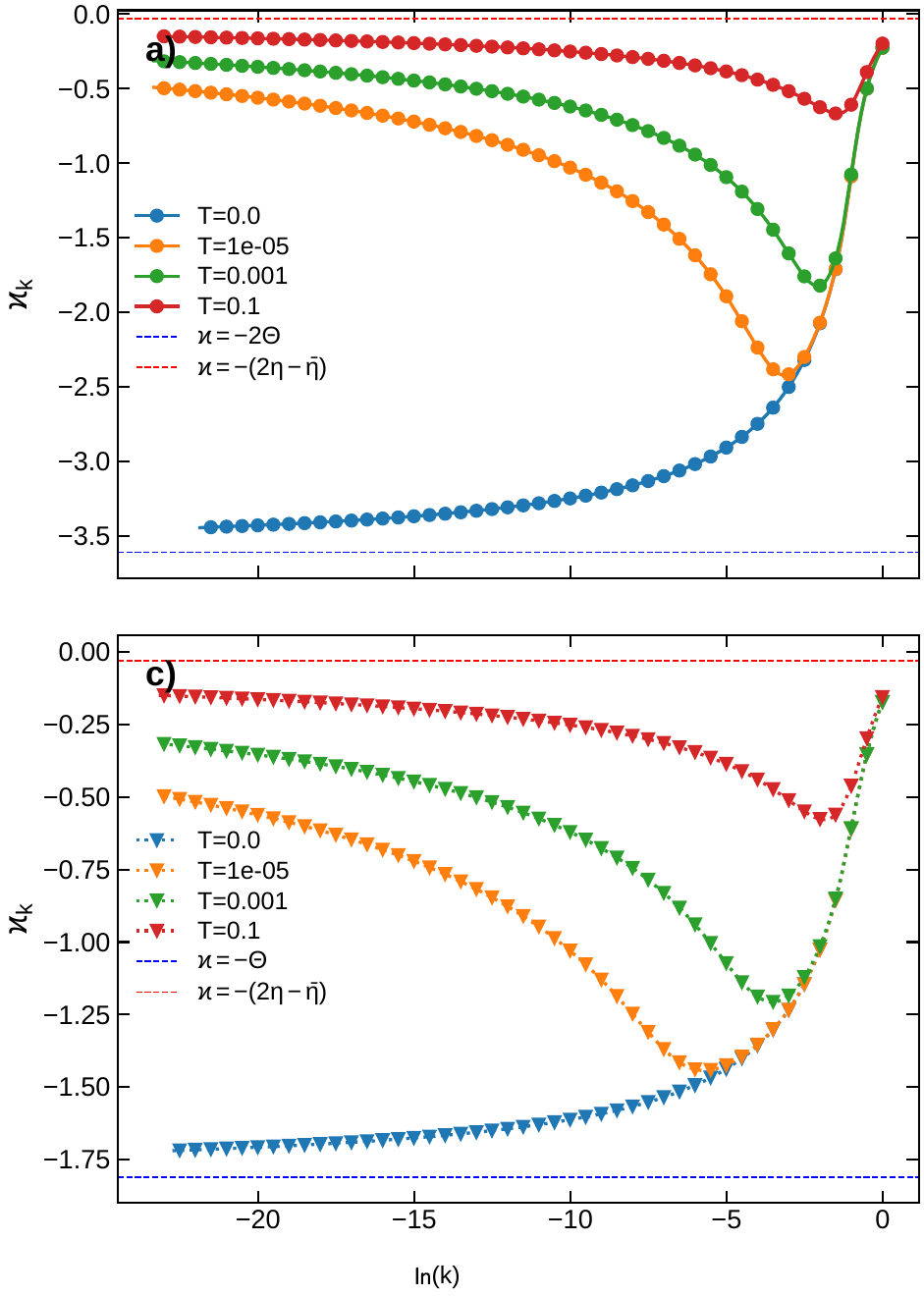}
	\includegraphics[width=0.49\textwidth]{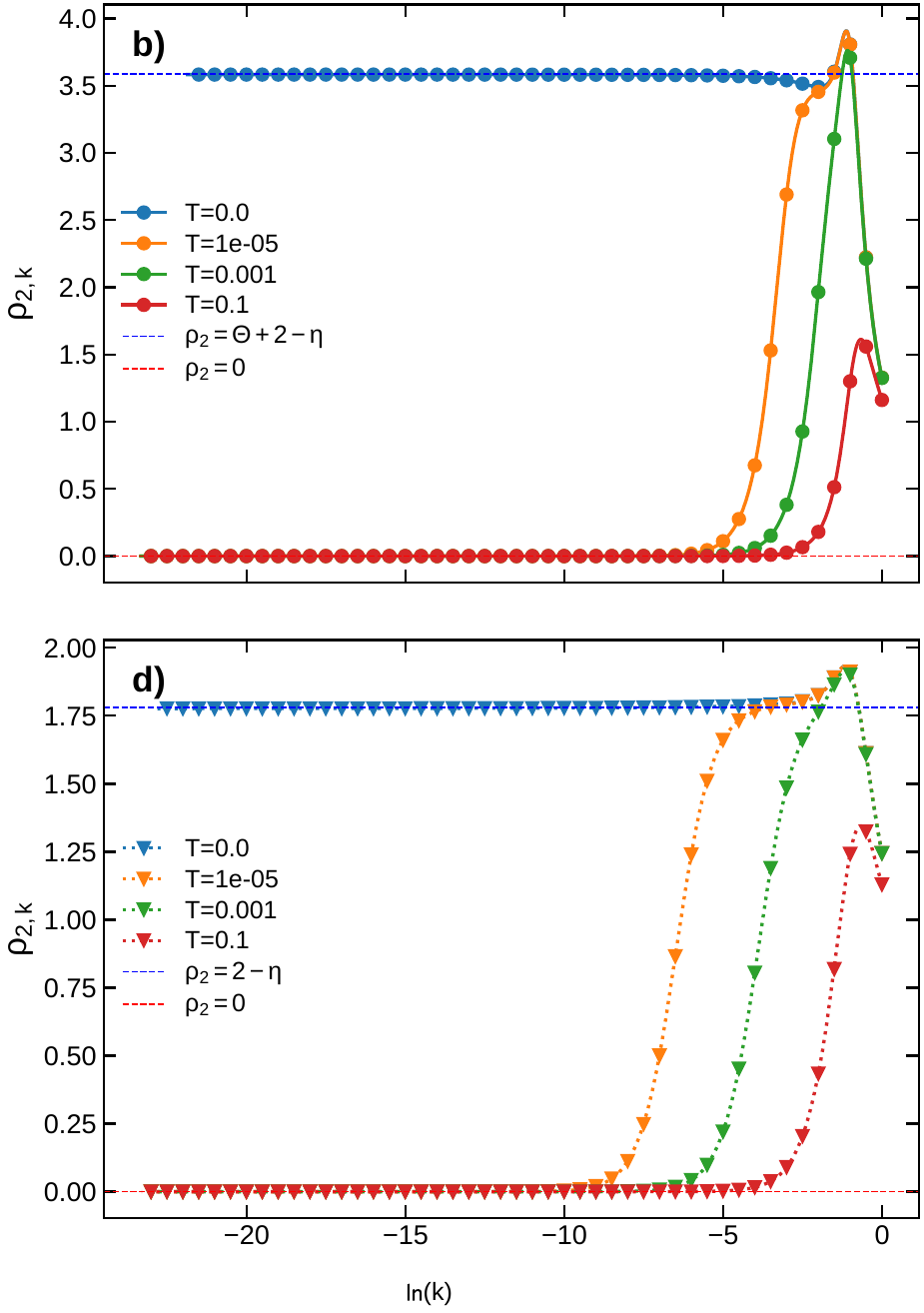}
	\caption{
		Flow of the effective exponents $\varkappa_k$ and $\rho_{2,k}$ as functions of the RG scale
		$\ln k$ for several values of the temperature $T$ in $d=4$.
		Panels (a) and (b)  show the flows associated with the initial condition $\propto \omega^2$, while panels (c) and (d) show the flow associated with the initial condition $\propto |\omega|$.  The horizontal dashed lines indicate the expected asymptotic values.
	}
	\label{fig:alphabeta_crossover_d4}
\end{figure}

To check the regularization mechanism independently we therefore also implement a self-consistent scheme in which $\varkappa$ is determined during the flow, analogously to the determination of $\eta$ in static NP-FRG calculations. In the large-$\bar{\omega}$ limit, Eq. \eqref{eq:fb_omega_large} becomes
\begin{equation}
	\label{eq:fb_omega_large_b}
	\partial_t f(\bar{\omega})\approx -\varkappa f(\bar{\omega})+\frac{\varkappa}{\sigma}\bar{\omega}f'(\bar{\omega})-c 
\end{equation}
with $c$ given by Eq. \eqref{eq:c_asymptotic}. Requiring the flow to vanish as $\bar{\omega}\to\infty$ yields a stable implicit equation for $\varkappa$, which is then inserted back into the flow of $\ln f$. Figure \ref{fig:alphabeta_crossover_d4} shows that this self-consistent procedure recovers $\varkappa=-\sigma\Theta$ and $\rho_2=(\sigma-1)\Theta+2-\eta$ at $T=0$, while after the quantum-to-classical crossover it gives $\varkappa=-(2\eta-\bar{\eta})$ and $\rho_2=0$ at finite temperature.

\begin{figure}[h!]
	\centering
	\includegraphics[width=0.65\textwidth]{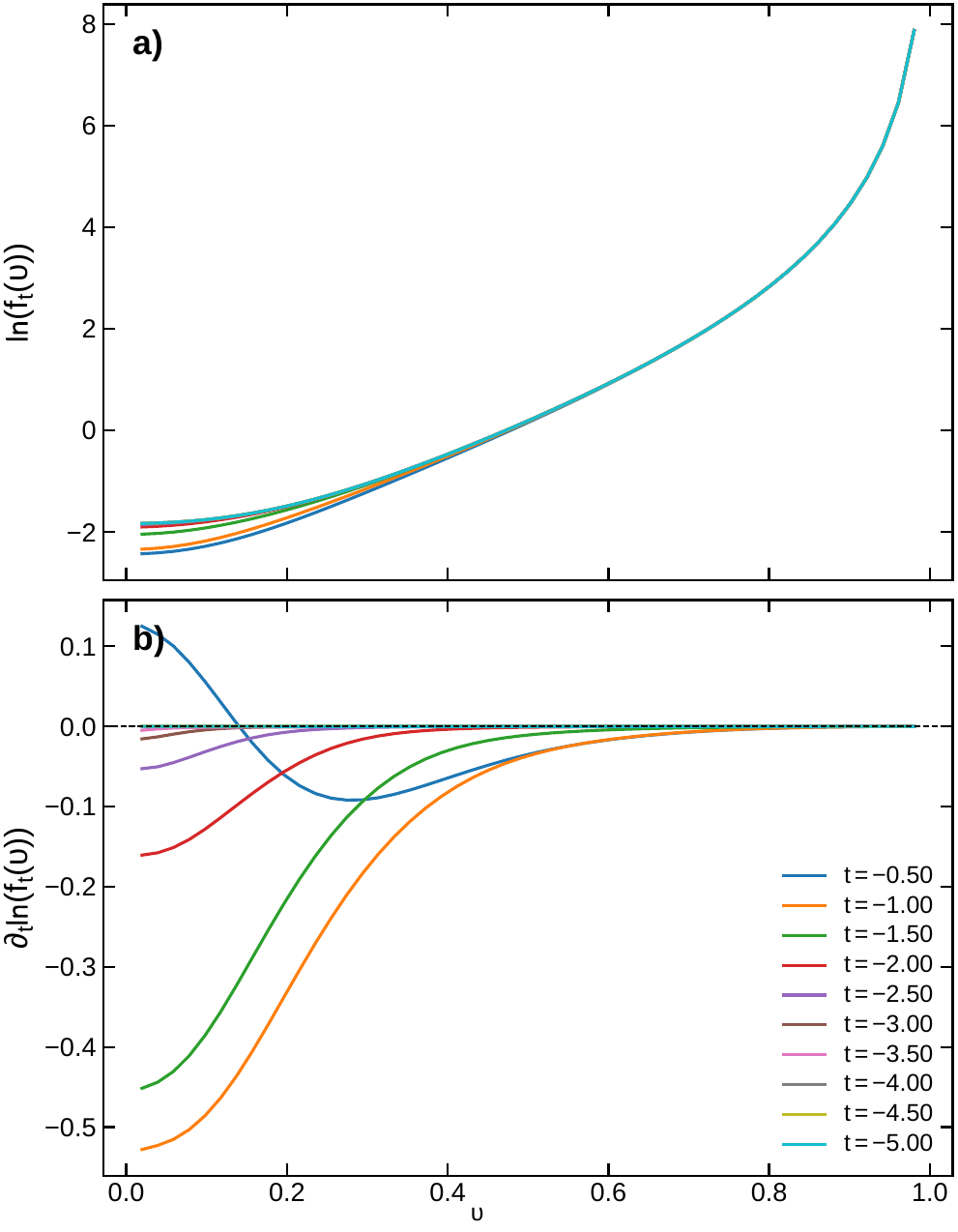}
	\caption{
		Flow of the $ln f_t(\upsilon)$ at $T=0$ and $d=4$ starting from an initial condition $\propto \omega^2$.
		Panel (a) shows the values and the panel (b)  shows the flows.
	}
	\label{fig:flow_f}
\end{figure}

Figure \ref{fig:flow_f} shows a representative flow of the dimensionless function $f(\bar{\omega})$. The function approaches a fixed shape in dimensionless frequency. We can easily see why this is the case by considering the Eq. \eqref{eq:fb_omega_large_b}. Inserting the form $f(\bar{\omega})=f_0+f_1\bar{\omega}^{\sigma}$ cancels the explicit $\bar{\omega}$ dependence, leaving only an equation for the constant $f_0$,
\begin{equation}
\partial_t f_0=-\varkappa f_0 -c.
\end{equation}
With the asymptotic solution and the regulator introduced above, $c$ can be expressed in terms of $f_0$ and $f_1$. The required coefficient $\bar{a}_2$ in Eq. \eqref{eq:c_asymptotic} is

\begin{equation}
	\label{eq:a2_asy0}
	\bar{a}_{2,asy.}\approx v_d\frac{c_1 (d\sqrt{f_1} +(d+\varkappa)\sqrt{f_0+c_1})\Gamma(d/2)}{4\sqrt{(f_0+c_1)f_1}(\sqrt{f_0+c_1}+\sqrt{f_1})^3},
\end{equation}
where $c_1$ is the amplitude of the frequency-dependent regulator and $d$ is the spatial dimension. This equation has a stable fixed point in $f_0$, consistent with the asymptotic limit of the functional flow shown in Fig. \ref{fig:flow_f}. 

\begin{figure}[h!]
	\centering
	\includegraphics[width=0.7\textwidth]{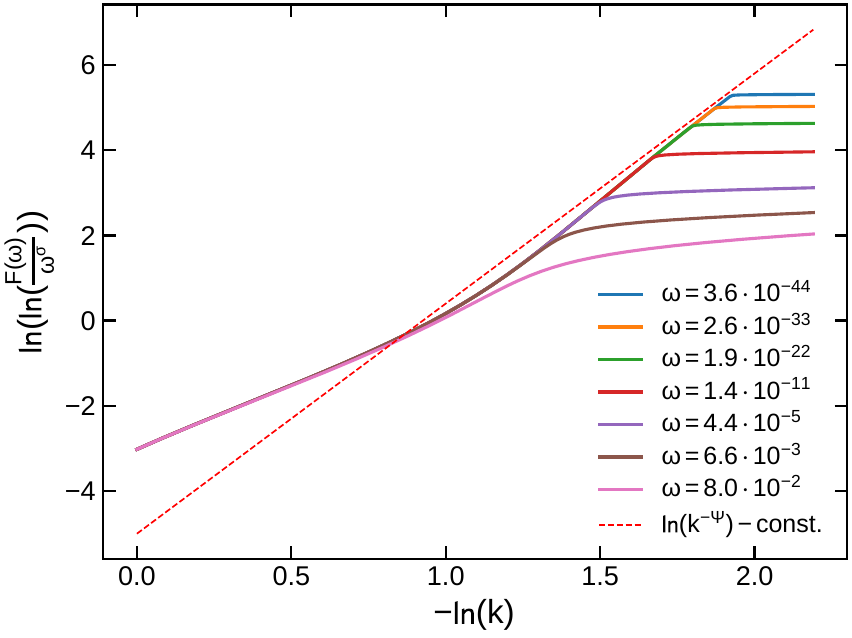}
	\caption{The figure shows the logarithm of the logarithm of the flow of the dynamical kernel for several dimensionful Matsubara frequencies, compared with the prediction from the activated scaling. The results shown are calculated at $T=0$ for $d=4$ with the initial condition for the dynamical kernel $\propto \omega^2$ in which case the activation exponent $\Psi\approx 5.4$ is predicted. The plateaus correspond to power law flow where $F\propto k^{\varkappa }$. Since the saturation happens at relatively small RG times even for minute frequencies, a slight deviation from the predicted exponent is still visible.}
	\label{fig:flow_matsubara}
\end{figure}

The dimensionless function $f(\bar{\omega})$ does not directly give the behavior at fixed dimensionful frequency: at fixed $\bar{\omega}$, the corresponding physical frequency is shifted toward the large-frequency sector during the flow. Nevertheless, $f$ determines $a_{2,k}$, which enters the flow of the dimensionful kernel $F(\omega)$. We therefore also integrate Eq. \eqref{eq:dyn_kernel2} at fixed dimensionful $\omega$ in parallel to $f(\bar{\omega})$. Figure \ref{fig:flow_matsubara} shows this flow over a broad frequency range in $d=4$. Even initial values as small as $\ln F(\omega_0)\approx -200$ (for $\omega\approx 3.6\cdot 10^{-44}$) reach the activated asymptotic regime within a short RG-time interval, after which the flow crosses over to the power-law regime once $F(\omega)\sim1$. 

The dimensionful flow equation is local in frequency: at fixed $\omega$, the flow of $F(\omega)$ depends only on $F(\omega)$ and on $a_{2,k}$. Thus the $T=0$ low-frequency behavior also informs the finite-temperature Matsubara grid. For sufficiently small $T$, the lowest modes $\omega_n=2\pi nT$ probe the same quantum activated regime before the quantum-to-classical crossover shown in Fig. \ref{fig:alphabeta_crossover_d4}. If the lowest mode still satisfies $F(\omega_1)\ll1$ after that crossover, the activation exponent crosses over to the classical value $\Psi=\Theta$ before eventually entering the power-law regime at $F(\omega_1)\sim1$. A full numerical analysis of these finite-temperature crossovers is left for future work.

		\subsection{Cautionary examples: apparent localization}
	\label{sec:cautionary_tales}
	
		We close this section by describing two calculations that produce a finite-scale divergence of the dynamical kernel, which could be mistaken for localization. 
		
		The first is a derivative expansion of the dynamical kernel. For simplicity, take $F(\omega)\approx F_1 \omega^2$ and extract the flow of $F_1$ in the regime $F(\omega)\ll1$. Equation \eqref{eq:dyn_kernel2} then gives
	\begin{eqnarray}
		\label{eq:dyn_kernelDE}
		\partial_t F_{1,k}&\approx&-F_{1,k}\frac{v_d}{2}\frac{a^*_1{\delta^*_1}^2}{a_{2,k}}k^{-\Theta} \int y^{d/2-1}dy e_0(y) g^3_s.
	\end{eqnarray}
We, however, have to evaluate $a_{2,k}$ consistently with the derivative expansion assumption. Keeping in mind that in the definition of the $a_{2,k}$, we arrive to a conclusion that 
	\begin{equation}
		a_{2,k}\propto \int d\omega \frac{a(\omega)}{(g^{-1}_s+k^{2-\eta} F_{1,k} \omega^2)^2} \propto \frac{1}{\sqrt{F_{1,k}}}.
	\end{equation}
 We stress that this statement holds true regardless of the regulator used and in particular even if we use the special regulator that we devised in the present work. (For general power $\sigma$ of frequency instead of $2$ we would have $(F_1)^{-1/\sigma}$). With this in mind we see that the equation for the coefficient $F_1$ is 
  \begin{equation}
 	\partial_t F_{1,k} \propto  -F^{1+p_1}_{1,k} k^{-p_2},  
 \end{equation}
where $p_1,p_2>0$. The solution of this equation is 
\begin{equation}
 	F_{1,k} =\left(p_1(\bar{F}_1-\frac{k^{-p_2}}{p_2})\right)^{\frac{1}{-p_1}},
 \end{equation}
where $\bar{F}_1$ is a positive constant given by the initial condition for $F_1$. This solution always diverges at a finite RG time given the conditions on exponents $p_1$ and $p_2$ and the fact that $F_1>0$ for all RG times.  This mechanism is closely related to the obstruction identified by  Gorokhov, Fisher, and Blatter in the FRG theory of quantum creep:
 a low-frequency truncation of the dynamical kernel as used in \cite{chauve00} would produce a spurious localization transition in a quantum problem, while a consistent treatment requires the full frequency-dependent dynamical spectrum \cite{Gorkhov02}.
 
Second scenario that we consider is the case when the full frequency dependence of the dynamical kernel is retained but the special frequency-regulating contribution - the second term in Eq.~\eqref{eq:a2k_def} - is not included.  In this case the flow still produces a divergence of the dynamical kernel at a finite RG time. This can be understood from a heuristic scaling argument (and seen trivially from the flow).
 
 Consider defining $a_{2,k}$ from Eq.~\eqref{eq:a2k_def} after omitting this second, special frequency-regulating term and keeping only the standard regulator (scaling like $k^{2-\eta}$). We then test the consistency of a putative scaling regime by inserting a trial exponent $\varkappa_0\leq 2-\eta$, following the discussion in Sec.~\ref{sec:criteria_consistency}. In this case the asymptotic scaling of $a_{2,k}$ is governed by the contribution \eqref{eq:nonregulating_term},
  \begin{equation}
 a_{2,k}\propto k^{\frac{\varkappa_0}{\sigma}-2(\varkappa_0-2+\eta)}.
 \end{equation}
Inserting this scaling back into Eq.~\eqref{eq:dyn_kernel_largew} gives, after one iteration, a new exponent
\begin{equation}
 \varkappa_1=-(4-\bar{\eta})+2\varkappa_0-\frac{\varkappa_0}{\sigma}.
\end{equation}

Given the critical exponents of the RFIM model in dimensions $2<d<6$ (see Table \ref{tab:exponent_exlamples}), it is not difficult to find that for either of the choices of $\sigma$ considered in the present article $4- \bar{\eta}>(1-\frac{1}{\sigma})(2-\eta)$, which implies that for all initial choices of $\varkappa_0\leq 2-\eta$  one finds $\varkappa_1<\varkappa_0$. The iteration therefore drives the exponent to progressively smaller values, corresponding to an increasingly strong growth of the dynamical kernel. In the numerical calculation this runaway manifests itself as a divergence at a finite RG time.
 
	
		\section{Summary of scaling results}
		\label{sec:summary}
        
\begin{figure}[h!]
	  	\centering
	  	\includegraphics[width=0.7\textwidth]{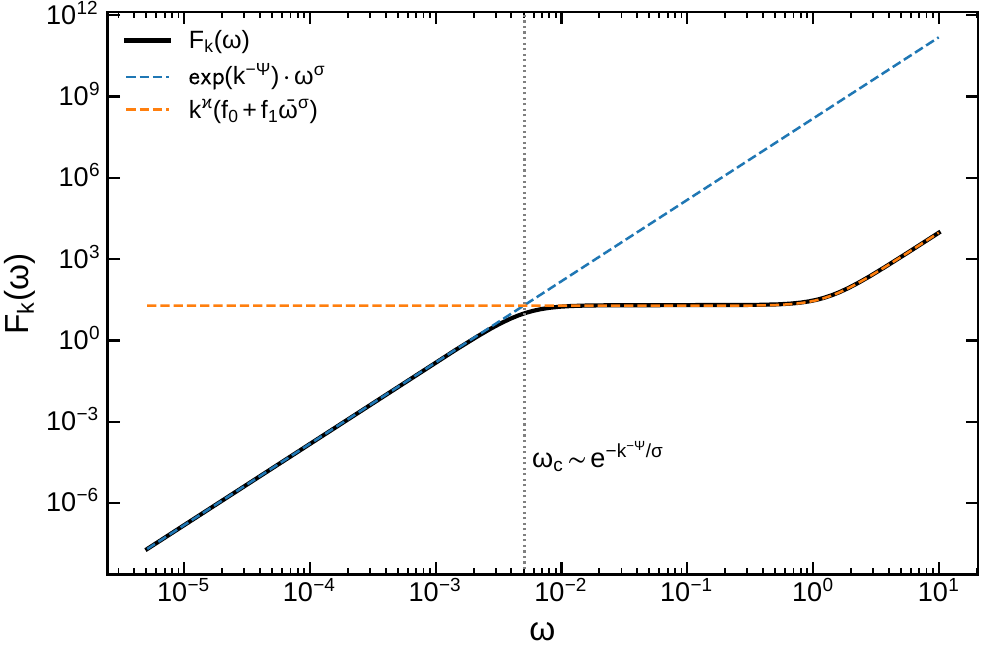}
	  	\caption{A schematic plot of the dynamical kernel in the asymptotic regime.	}
	  	\label{fig:dyn_kernel_schematic}
	  \end{figure}
	  
		  We summarize the asymptotic shape of the zero-temperature dynamical kernel in Fig. \ref{fig:dyn_kernel_schematic}. For a bare initial condition $F\propto \omega^\sigma$, the scaling form is
	  	  \begin{equation}
	  	F_k(\omega) =
	  	\begin{cases}
	  		const. \cdot\exp(k^{-\Psi})\,\omega^\sigma, 
	  		& F(\omega) \lesssim 1, \\[4pt]
	  		k^\varkappa\left(f_0+f_1\bar{\omega}^{\sigma}\right), 
	  		& F(\omega) \gtrsim 1 .
	  	\end{cases}
	  \end{equation}

	  	\begin{figure}[h!]
	  	\centering
	  	\includegraphics[width=0.65\textwidth]{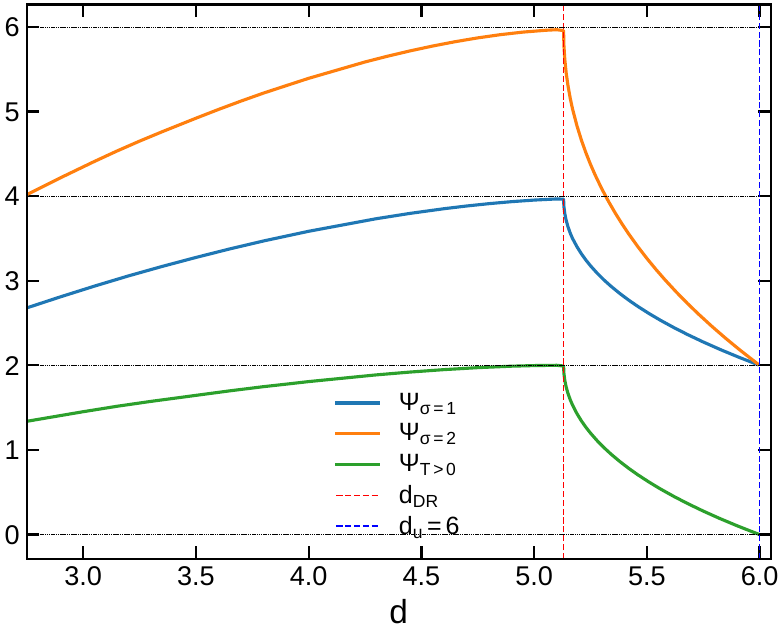}
	  	\caption{The activation exponents for different cases.	Exponent $\sigma$ denotes the power of the frequency in the dynamical kernel in the UV. They are given by the static critical exponents of the equilibrium random field Ising model which are calculated with the exponential regulator (see Eq. \eqref{eq:regulator_def}) with the prefactor $c_0=1.44$ (for the explanation of this number see e.g. caption of the Tab. \ref{tab:exponent_exlamples} or \cite{tarjus25}). The thin dashed lines are the values of the exponents in the limit $d\to d_{DR}$, the dimensional reduction breaking dimension. }
	  	\label{fig:psi_vs_d}
	  \end{figure}
	  
	  \begin{figure}[h!]
	  	\centering
	  	\includegraphics[width=0.65\textwidth]{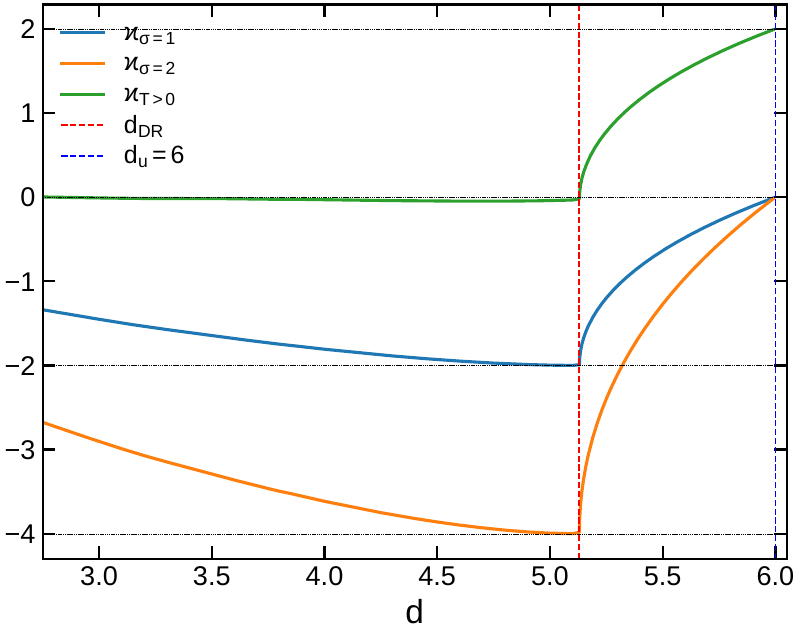}
	  	\caption{The scaling exponents of the dynamical kernel $\varkappa$ for different cases.	Exponent $\sigma$ denotes the power of the frequency in the dynamical kernel in the UV.  They are given by the static critical exponents of the equilibrium random field Ising model which are calculated with the exponential regulator (see Eq. \eqref{eq:regulator_def}) with the prefactor $c_0=1.44$. The exponent $\varkappa_{T>0}$ for $d<d_{DR}$ is always very small but negative number as the Schwartz-Soffer inequality $2\eta-\etab>0$ \cite{schwartz85} is respected by the calculations \cite{tarjus13a}. For $d>d_{DR}$ however it turns positive which does not make the calculations inconsistent since $\Psi_{T>0}>0$, i.e. $\varkappa<2-\eta$ as explained in the Sec. \ref{sec:criteria_consistency}. }
	  	\label{fig:alpha_vs_d}
	  \end{figure}	
	  
		  The asymptotic exponents are determined by the static RFIM exponents and by the UV frequency power $\sigma$. 
		  
		  We now extend the result to $d>d_{DR}\approx 5.1$. In the classical RFIM, the barrier exponent becomes $\Psi_{d>d_{DR}}=\Theta-2\lambda$, where $\lambda$ is the cusp subdominance exponent related to the number of critical avalanches \cite{tarjus13,tarjus25}, and $\Theta_{d>d_{DR}}=2$ \cite{balog_activated}. The exponent $\lambda$ is a property of the classical RFIM fixed point. In the present calculation the same modification enters because the scaling function $\Delta^{(0,1)}_k$ in Eq. \eqref{eq:Delta_1_scaling} acquires an additional scaling factor $k^{\lambda}$ (with $\lambda>0$) for $d>d_{DR}$ as the cusp flows to $0$ in the $k\to 0$ limit.  
	  
		  The large-frequency scaling equation \eqref{eq:dyn_kernel_largew} is therefore modified, giving
	 \begin{equation}
	  	\label{eq:eq:alphabeta_dgtdr}
	  	\varkappa_{d>d_{DR}}=-(\rho_{2,d>d_{DR}}+2\eta-\bar{\eta})+2\lambda.
	  \end{equation}
	    The self-consistency condition for $\rho_{2,d>d_{DR}}$ follows from the scaling of the regulating term in $a_{2,k}$, Eq. \eqref{eq:regulating_term}:
	  \begin{equation}
	  	\rho_{2,d>d_{DR}}=(\sigma-1)(\Theta-2\lambda)+(2-\eta).
	  \end{equation}
	  Inserting this into Eq. \eqref{eq:eq:alphabeta_dgtdr} yields
	  \begin{equation}
	  \varkappa_{d>d_{DR}}=-\sigma(\Theta-2\lambda).
	  \end{equation}
	    The zero-temperature activation exponent is $\Psi_{d>d_{DR}}=\rho_2+\Theta-2\lambda$, namely
	  \begin{equation}
	  \Psi_{d>d_{DR}}=\sigma(\Theta-2\lambda)+(2-\eta).
	  \end{equation}
	  The consistency criterion of Sec.~\ref{sec:criteria_consistency} becomes $\rho_2+\Theta-2\lambda>0$, or equivalently
	  \begin{equation}
	  	\label{eq:consistency_criterion_d>dr}
	  	\Psi_{d>d_{DR}}>0,
	  \end{equation}
	  confirming the generality of Eq.~\eqref{eq:validity_criterion_+}.
	  
	The barrier exponent $\Psi$ and the dynamical-kernel scaling exponent $\varkappa$ are therefore
	  \begin{equation}
		  	\label{eq:psi_all}
		\Psi =
		\begin{cases}
			\sigma\Theta+2-\eta, 
			& T=0, d< d_{DR}, \\[4pt]
			\sigma(\Theta-2\lambda)+2-\eta, 
			& T=0 , d\geq d_{DR},\\[4pt]
			\Theta, 
			& T>0, d< d_{DR} , \\[4pt]
			\Theta-2\lambda, 
			& T>0 , d\geq d_{DR},
		\end{cases}
	\end{equation}
	  where $\Theta=2-\bar{\eta}+\eta$ (remember that $\eta=\bar{\eta}$ for $d>d_{DR}$) and
	  \begin{equation}
		  	\label{eq:alpha_all}
		\varkappa =
		\begin{cases}
			-\sigma\Theta, 
			& T=0, d<d_{DR}, \\[4pt]
			-\sigma(\Theta-2\lambda), 
			& T=0, d\geq d_{DR} , \\[4pt]
			-(2\eta-\bar{\eta}), 
			& T>0 , d<d_{DR}, \\[4pt]
			-\eta+2\lambda, 
			& T>0, d\geq d_{DR} .
		\end{cases}
	\end{equation}

		Figures \ref{fig:psi_vs_d} and \ref{fig:alpha_vs_d} show these exponents over the numerically accessible range $2.75<d<6$, using $\lambda$ from Ref. \cite{tarjus25}. Representative values are given in Table \ref{tab:exponent_exlamples}. The finite-temperature activation exponent vanishes as $d\to6$, whereas the zero-temperature quantum exponents approach $\Psi_\sigma\to2$ for all $\sigma$. In all cases shown, $\Psi>0$, satisfying the consistency criterion of the calculation.

\section{Conclusions and outlook}	
\label{sec:conclusion}

We have provided, to our knowledge, the first NP-FRG calculation of the full frequency-dependent dynamic kernel in the quantum random-field Ising model. The resulting flow is consistent with activated dynamics exhibiting two distinct scaling regimes: the first is due to quantum tunneling at zero temperature, and the second, at finite temperature, is asymptotically due to thermal activation (after a crossover in which the quantum regime is felt (see Fig. \ref{fig:alphabeta_crossover_d4})). The extremely slow dynamics at  $T=0$  is especially subtle because direct implementations of the flow can generate divergences in the dynamical sector at finite length scales, which may be misinterpreted as an apparent localization of the dynamics~\cite{Gorkhov02,grison26}.Within the present calculation, this difficulty is resolved by retaining the full frequency dependence of the dynamic kernel and by using a frequency-dependent regulator adapted to its running scale. This regulator construction is, to our knowledge, new. We do not claim that it is the only possible way of regularizing the flow, but it provides a self-consistent scheme in which the infrared dynamics remains controlled and the apparent localization is replaced by activated scaling.

\begin{table}[th!]
		\centering
		\begin{tabular}{l|l |l| l | l | l | l | l | l | l}
			\hline
			$d$ & $\eta$ & $\bar{\eta}$ &$\Theta(=\Psi_{T>0}$) & $\Psi_{T=0,\sigma=1}$  & $\Psi_{T=0,\sigma=2}$ & $\varkappa_{T>0}$ & $\varkappa_{T=0,\sigma=1}$   & $\varkappa_{T=0,\sigma=2}$\\
			\hline
			$3$ & $0.56$ & $1.11$  & $1.45$ & $2.89$  & $4.35$ & $-0.009$ & $-1.45$  & $-2.90$  \\
			$4$ & $0.22$ & $0.42$  & $1.81$ & $3.58$  & $5.39$ & $-0.03$  & $-1.81$ & $-3.91$   \\
			$5$ & $0.044$ & $0.048$ & $1.997$ & $3.952$ & $5.949$ &  $-0.004$  & $-1.997$& $-3.993$  \\
			\hline
		\end{tabular}
		\caption{Critical exponents relevant for describing the behavior of the dynamical kernel for several spatial dimensions. All the values were calculated with the exponential regulator optimized slightly above $d_{DR}$ (its prefactor is $c_0=1.44$ following Eq. \eqref{eq:regulator_def}). Absolute error bars are not available within the present calculation (see e.g. \cite{depolsi20} for a discussion). One can get the feeling for them noting that the exponent $\eta$ at $d=5$, since the breaking of the dimensional reduction property is minimal \cite{balog19a}, should compare closely (within a percent) of the best known estimate of the $\eta$ for the pure Ising model at $d=3$, which is $\eta\approx 0.03629761(5)$ found by conformal bootstrap method \cite{chang25},. Note as well that all numerical determinations of the critical exponents $\eta$ and $\bar{\eta}$ for the RFIM at $d$ represented in the table are roughly within $15\%$ of the values we quote here \cite{fytas18}, being marginally more consistent as the dimension decreases. } 
		\label{tab:exponent_exlamples}
	\end{table}

The mechanism uncovered here may have implications beyond the quantum RFIM. A common feature of several disorder-dominated quantum problems is the appearance of a nonanalytic cusp in the renormalized disorder cumulant, together with a fluctuation-induced boundary layer that rounds this cusp at any nonzero running scale. This structure occurs, for example, in pinned elastic manifolds \cite{chauve01,ledoussal08,wiese22,ledoussal02_dep} and in functional RG descriptions of Bose- and Mott-glass phases \cite{daviet20,dupuis24,grison26}. In such systems the cusp encodes the presence of many competing metastable configurations or droplets, while its rounding describes the residual coupling between these configurations induced by thermal or quantum fluctuations.

Our results suggest that this structure should be treated with particular care in dynamical calculations. A finite-scale singularity in a truncated dynamical sector need not by itself imply a true localization of the dynamics. In the QRFIM, keeping the full frequency dependence and adapting the regulator to the running dynamical kernel converts the apparent divergence into activated scaling. It is therefore natural to ask whether analogous mechanisms operate in equilibrium random elastic manifolds, quantum creep problems, and disordered one-dimensional Bose fluids, where related cusp and boundary-layer structures are already known to play a central role.

The equilibrium problem studied here is also a necessary first step toward driven systems, such as quantum creep in random elastic media \cite{Gorkhov02}. In the driven case, the equilibrium exponents controlling the rescaling of the dynamical kernel and the typical barrier height remain essential inputs, but the physical motion depends additionally on the applied drive. At finite temperature the creep motion proceeds by thermal activation over scale-dependent barriers, whereas at $T=0$ it proceeds by quantum tunneling through such barriers. Extending the present NP-FRG framework to this driven setting should therefore make it possible to connect the equilibrium activated dynamics derived here with the force-dependent quantum creep laws of pinned manifolds.

In summary, our calculation supports a coherent picture in which the apparent localization found in insufficiently resolved flows is replaced by activated dynamics once the full frequency dependence of the dynamical kernel is treated consistently. The conclusion relies on the regulator construction introduced here, and it will be important to test its robustness against alternative regularization schemes and improved truncations. Nevertheless, the underlying physical mechanism is not tied to microscopic details of the QRFIM: it follows from general structural properties of disorder-dominated fixed points, namely anomalous cumulant scaling, cusp formation, and fluctuation-induced cusp rounding. This opens a route toward a unified NP-FRG treatment of activated dynamics and quantum tunneling in random-field systems, random elastic manifolds, Bose-glass problems, and driven disordered quantum matter.

\section*{Acknowledgements}
IB wishes to thank the Physics Laboratory (LPENSL) of the École normale supérieure de Lyon where a part of this work was done. We wish to thank Gilles Tarjus and Nicolas Dupuis for useful comments during the work on this problem. 

\paragraph{Author contributions}
IB and AF conceived the research. IB and AF contributed to writing the paper. IB ans LŠ performed the calculations and simulations.

\paragraph{Funding information}
IB and LŠ are supported by the Croatian Science fund project HRZZ-IP-10-2022-9423 (I.Balog). IB, LŠ and AF are funded by the CNRS IRP project DisQ. IB wishes to acknowledge the support of the INFaR and FrustKor
projects financed by the EU through the National Recovery and Resilience Plan (NRRP) 2021-2026 as well as the QuantiXLie Centre of Excellence , a project cofinanced by the Croatian Government and European Union through the European Regional Development Fund-the Competitiveness and Cohesion Operational Programme (Grant No. KK.01.1.1.01.0004).

\newpage
\appendix

\section{Quantum corrections to the one-copy sector}
\label{sec:appx_1copy_corrections}

The quantum corrections to the flows of $U''$ and $Z$ are given by the following expressions 
\begin{eqnarray*}
	\beta'_{U''} &=&\int_q\int_\omega \dfrac{\partial_t R_1(q^2,\omega)}{2}\hat{G}[\phi;q^2,\omega]^2\Big(-q^2 Z^{(2)}(\phi)-U^{(4)}(\phi)+2\hat{G}[\phi;q^2,\omega] \\
 && \times\Big(q^2 Z'(\phi)+U^{(3)}(\phi)\Big)\Big),
\\
	\beta'_{Z}&=&\int_q\int_\omega \dfrac{\hat{G}[\phi;q^2,\omega]}{2d} \Big(\hat{G}[\phi;q^2,\omega]\Big(q^2 Z'(\phi)+U^{(3)}(\phi)\Big)\Big(\hat{G}[\phi;q^2,\omega]\Big(2q^2 \partial_t R_1^{(2,0)}(q^2,\omega)\\
&&  \times \Big(q^2 Z'(\phi)+ U^{(3)}(\phi)\Big)+ \partial_t R_1^{(1,0)}(q^2,\omega)\Big((4+d)q^2 Z'(\phi)+d U^{(3)}(\phi)\Big) \Big)
	\\
&&+8q^2  \partial_t R_1^{(2,0)}(q^2,\omega)\Big(q^2 Z'(\phi)+U^{(3)}(\phi)\Big)\hat{G}^{(0,1,0)}[\phi,q^2,\omega]\Big)
	\\
&&+ \partial_t R_1(q^2,\omega)\Big(2\hat{G}[\phi;q^2,\omega]Z'(\phi)\Big((1+2d)q^2Z'(\phi)+2d U^{(3)}(\phi)\Big)\\
&& +4q^2\Big(q^2Z'(\phi)+U^{(3)}(\phi)\Big)^2\hat{G}^{(0,1,0)}[\phi,q^2,\omega]^2+\hat{G}[\phi;q^2,\omega]\Big(-d Z^{(2)}(\phi)
	\\
&&+3\Big(q^2 Z'(\phi)+U^{(3)}(\phi)\Big)\Big(\Big((4+d)q^2 Z'(\phi)+d U^{(3)}(\phi)\Big)\hat{G}^{(0,1,0)}[\phi,q^2,\omega]
	\\
&&+2q^2\Big(q^2Z'(\phi)+U^{(3)}(\phi)\Big)\hat{G}^{(0,2,0)}[\phi,q^2,\omega]\Big)\Big)\Big)\Big).
\end{eqnarray*}
These equations coincide with the corresponding pure quantum Ising flow equations. All contributions contain full Matsubara integrals and are suppressed asymptotically once the dynamical kernel becomes large. The least suppressed terms contain the fewest propagators. For the dominant contribution to the flow of $U''$, we use the asymptotic form $F(\bar{\omega})=k^{\varkappa}(f_0+f_1 \bar{\omega}^{\sigma})$ and keep the leading frequency-regulator contribution, obtaining
\begin{equation}
\int_q\int_{\omega} \partial_t R_{1}(q^2,\omega)\hat{G}[\phi;q^2,\omega]^2U^{(4)}(\phi)\approx k^{\varkappa\left(\frac{1-\sigma}{\sigma}\right)+\Theta+4} \int y^{d/2-1}dy\int d\bar{\omega} \frac{e_1(y,\bar{\omega}^2)}{(f_0+c_1 +f_1 \bar{\omega}^\sigma )^2}u^{(4)}(\varphi).
\end{equation}
Given the expression for $\varkappa$ in the $T=0$ cases we find that such a term is subdominant if 
\begin{equation}
\label{eq:1copy_subdominance}
   \sigma\Theta +4>2-\eta,
\end{equation}
where the $2-\eta$ is the scaling power of $U''$ coming from the $T=0$ contributions. This inequality appears always to be satisfied for the RFIM (see Table \ref{tab:exponent_exlamples}). Contributions in the flow of $U''$ with $\hat{G}^3$ are even more suppressed by an additional factor of $k^{-\varkappa+2-\eta}$.

The most relevant corrections to the field renormalization $Z$ obey the same scaling logic and lead to the same condition \eqref{eq:1copy_subdominance}.

\section{Corrections to the flow of the dynamical kernel}
\label{sec:appx_F_corrections}

The full flow of the dynamical kernel can be written as
\begin{equation}
	\label{eq:flow_of_F_full}
	\partial_t F(w) = (I_0(w)+I_1(w)+I_2(w))-(I_0(0)+I_1(0)+I_2(0)),
\end{equation}
where 
\begin{eqnarray}
	I_0(w) &=& -\int_q \dfrac{\partial_t R_1(q^2,0)}{4}\hat{G}[0;q^2,w]^2\Delta^{(0,2)}(0,0)\\
	I_1(w)&=&\int_q \partial_t R_2(q^2,0)\hat{G}[0;q^2,0]^2\hat{G}[0;q^2,w]\Big(q^2 Z'(0)+U^{(3)}(0)\Big)^2\nonumber\\
	&+&\int_q \dfrac{\partial_t R_1(q^2,0)}{4}\hat{G}[0;q^2,w]\Bigg(4\Big(\Delta(0,0)-R_2(q^2,0)\Big)\hat{G}[0;q^2,0]^2
    \nonumber\\
	&\times&\Big(2\hat{G}[0;q^2,0]+\hat{G}[0;q^2,w]\Big)\Big(q^2 Z'(\phi)+U^{(3)}(\phi)\Big)^2\nonumber\\	
    &-&4\hat{G}[0;q^2,0]\Big(\hat{G}[0;q^2,0]+\hat{G}[0;q^2,w]\Big)\Big(q^2 Z'(0)+U^{(3)}(0)\Big)\Delta^{(1,0)}(0,0)\nonumber\\
    &+&\hat{G}[0;q^2,w]\Delta^{(2,0)}(0,0)\Bigg)\\
	I_2(w) &=& \int_q\int_\omega \dfrac{\partial_t R_1(q^2,\omega)}{2}\hat{G}[0;q^2,\omega]\hat{G}[0;q^2,\omega+w]\Big(\hat{G}[0;q^2,\omega]+\hat{G}[0;q^2,\omega+w]\Big)\nonumber\\
	&\times&\Big(q^2 Z'(0)+U^{(3)}(0)\Big)^2.
\end{eqnarray}
The subtraction ensures that the flow vanishes at $w=0$. The leading contribution is $I_0$, which gives Eq. \eqref{eq:dyn_kernel0}.

The remaining terms define two correction classes, $I_1$ and $I_2$. 

The $I_1$ class is represented by pairs of terms such as
\begin{equation}
	\label{eq:F_class_1_example}
	\frac{1}{4}\int_q\partial_t R_{1}(q^2,0)\Delta^{(2,0)}(0,0)(\hat{G}[0;q^2,w]^2-\hat{G}[0;q^2,0]^2),
\end{equation}
All terms in this class contain two-copy propagators or vertices, so their internal Matsubara frequency is fixed to zero.

The magnitude of this class can be assessed from Eq. \eqref{eq:F_class_1_example}. All $I_1$ terms scale in the same way, because the estimate does not depend on the dynamical-kernel power. In the large dimensionful external-frequency limit, the frequency-dependent part vanishes and the contribution scales as
 \begin{equation}
 	-\int_q\partial_t R_{1}(q^2,0)\Delta^{(2,0)}(0,0)\hat{G}[0;q^2,0]^2\propto k^{2-\eta}.
 \end{equation}
Such terms scale as expected from the second derivatives of the first cumulant of the action. 

The $I_2$ class contains no two-copy propagator or disorder vertex, and therefore includes a full Matsubara integral. As for the corrections to $U''$ and $Z$, these terms are suppressed by powers of the propagator. At the present truncation level they also vanish by symmetry at $\phi=0$, because they are proportional to $(\Gamma^{(3)}_{1,k})^2$. 


\section{Frequency-regulator diagnostics}
\label{sec:appx_regulator_insights}

Because the frequency regulator used here is nonstandard, we record several diagnostics that guided its construction. 

First of all we note that the regulating term of the $a_{2,k}$ (see Eq. \eqref{eq:regulating_term})  should be positive. If it were not, it would necessarily  lead to $a_{2,k}$ going to $0$ at a finite RG time since it scales dominantly to the other nonregulating term which is indeed always positive (see Eq. \eqref{eq:nonregulating_term}) and thus leading to a divergence in the dynamical kernel. An obvious question then is why it would be negative (and why would the other term generically be positive)? The fundamental reason  follows from the inverted logic of the characteristic frequencies compared with the characteristic momentum, which is $k$. 

As we noted previously, the characteristic frequency $\omega_0\propto k^{\frac{\varkappa}{\sigma}}$, flows to infinity in the RG flow. Consider the two expressions for the dimensionless regulator time derivatives $e_0$ and $e_1$ shown in Equations \eqref{eq:e0_definition} and \eqref{eq:e1_definition}. We thus see that the dimensionless time derivative of the standard regulating term ($e_0$ - Eq.  \eqref{eq:e0_definition} ) is generically positive since the first (and the largest) contribution is multiplied by a positive exponent $2-\eta$. The dimensionless time derivative of  the special regulating term ($e_1$ - Eq. \eqref{eq:e1_definition})  has the first contribution multiplied by a negative exponent $\varkappa$. What is quite easily seen is that the first contribution to $e_1$ is largest at small dimensionless frequencies, while other two contributions are indeed positive but dominate at large dimensionless frequencies. 

The conclusion is that if we want the regulating term of the $a_{2,k}$ positive we should allow the positive terms to take over by not introducing more cutting than is necessary for all the relevant Matsubara integrals to converge  in the dimensionless frequency sector.  This conclusion is consistent with the conclusion from \cite{Gorkhov02}, where they find that the large frequency region regulates the flow of the dynamical kernel as well. 

Let us give a few examples of this behavior. We wish to calculate the expression $a_{2,k}$ in the asymptotic limit when $F(\bar{\omega})=k^{\varkappa}f(\bar{\omega})>>1$.  We assume the shape $f(\bar{\omega})\approx f_0 + f_1 \bar{\omega}^\sigma$. The expression for the shape of the regulator that we used in the work is already given by the Eq. \eqref{eq:a2_asy0} for the case $\sigma=2$. In order for this to be positive, one must have $c_1 (d\sqrt{f_1} +(d+\varkappa)\sqrt{f_0+c_1})>0$. The only possible source of the problem might be $d+\varkappa$, however it appears that the RFIM critical exponents (see Table \ref{tab:exponent_exlamples}) always conspire to give $d+\varkappa>0$. 

Imagine next the possibility where we would cut exponentially in terms of the dimensionless frequencies and use instead of Eq. \eqref{eq:s1regulator} the following
\begin{equation}
	\label{eq:s1regulator_cut}
	s_{1,alt}(\frac{q^2}{k^2},\frac{\omega^2}{\omega_{0}^2})=\frac{\frac{\omega^2}{\omega_{0}^2}}{1+\frac{\omega^2}{\omega_{0}^2}}\cdot s_0(\frac{q^2}{k^2}+\frac{\omega^2}{\omega_{0}^2}). 
\end{equation}
Assuming the same asymptotic shape of the dynamical kernel function and $\sigma=2$ we get
\begin{equation}
	\bar{a}_{2,asy,alt}\approx \int_{\bar{\omega}}\frac{e^{-\bar{\omega}^2} c_1\bar{\omega}^2
		\left[
		d + (d+2\varkappa)\bar{\omega}^2 + \varkappa\bar{\omega}^4
		\right]
		\Gamma\!\left(\frac{d}{2}\right)
	}{
		(1+\bar{\omega}^2)^2
		(f_0+c_1+f_1\bar{\omega}^2)^2
	}.
\end{equation}
We show the integrand in $\bar{\omega}$ because the integrated expression is quite complicated but we find that it is very difficult to find a combination of $c_1$, $f_0$ and $f_1$ positive where this integral is positive with $\varkappa$ that is physical. 

We even went a step further and tried the regulator dependence 
\begin{equation}
	\label{eq:s1regulator_+}
	s_{1,alt+}(\frac{q^2}{k^2},\frac{\omega^2}{\omega_{0}^2})=\frac{\left(\frac{\omega^2}{\omega_{0}^2}\right)^{1+r/2}}{1+\frac{\omega^2}{\omega_{0}^2}}\cdot s_0(\frac{q^2}{k^2}),
\end{equation}
with the exponent $r>0$ but such that the Matsubara integral in Eq. \eqref{eq:regulating_term} converges for a given $\sigma$ on the infinite domain of $\bar{\omega}$. Note that such a regulator diverges at $\bar{\omega}\to \infty$. Our preliminary calculations show that such choices work just as well as the regulator that we used in the present work (Eq. \eqref{eq:s1regulator}) and give the same asymptotic behavior. It however ensures the positivity of the regulating contribution of $a_{2,k}$ (see Eq. \eqref{eq:regulating_term}) even more strongly than the choice \eqref{eq:s1regulator} we used throughout the work.
	
	\bibliographystyle{apsrev4-1}
	\bibliography{references}

@article{MicnasChao1984,
  title = {Quantum critical behavior in the presence of a random field},
  author = {Micnas, R. and Chao, K. A.},
  journal = {Phys. Rev. B},
  volume = {30},
  issue = {11},
  pages = {6785(R)--6787(R)},
  numpages = {0},
  year = {1984},
  month = {Dec},
  publisher = {American Physical Society},
  doi = {10.1103/PhysRevB.30.6785},
  url = {https://link.aps.org/doi/10.1103/PhysRevB.30.6785}
}

@article{silevitch2007,
  title   = {A ferromagnet in a continuously tunable random field},
  author  = {Silevitch, D. M. and Bitko, D. and Brooke, J. and Ghosh, S. and Aeppli, G. and Rosenbaum, T. F.},
  journal = {Nature},
  volume  = {448},
  number  = {7153},
  pages   = {567--570},
  year    = {2007},
  doi     = {10.1038/nature06050},
  pmid    = {17671498}
}

@article{Motrunich2000,
  title = {Infinite-randomness quantum Ising critical fixed points},
  author = {Motrunich, Olexei and Mau, Siun-Chuon and Huse, David A. and Fisher, Daniel S.},
  journal = {Phys. Rev. B},
  volume = {61},
  issue = {2},
  pages = {1160--1172},
  numpages = {0},
  year = {2000},
  month = {Jan},
  publisher = {American Physical Society},
  doi = {10.1103/PhysRevB.61.1160},
  url = {https://link.aps.org/doi/10.1103/PhysRevB.61.1160}
}

@article{senthil98,
	author  = {Senthil, T.},
	title   = {Properties of the random-field {I}sing model in a transverse magnetic field},
	journal = {Phys. Rev. B},
	volume  = {57},
	pages   = {8375},
	year    = {1998},
	doi     = {10.1103/PhysRevB.57.8375}
}

@article{fisher92,
	author  = {Fisher, Daniel S.},
	title   = {Random transverse field {I}sing spin chains},
	journal = {Phys. Rev. Lett.},
	volume  = {69},
	pages   = {534--537},
	year    = {1992},
	doi     = {10.1103/PhysRevLett.69.534}
}

@article{fisher95,
	author  = {Fisher, Daniel S.},
	title   = {Critical behavior of random transverse-field {I}sing spin chains},
	journal = {Phys. Rev. B},
	volume  = {51},
	pages   = {6411--6461},
	year    = {1995},
	doi     = {10.1103/PhysRevB.51.6411}
}

@article{vojta06,
	author  = {Vojta, Thomas},
	title   = {Rare region effects at classical, quantum, and nonequilibrium phase transitions},
	journal = {J. Phys. A: Math. Gen.},
	volume  = {39},
	number  = {22},
	pages   = {R143--R205},
	year    = {2006},
	doi     = {10.1088/0305-4470/39/22/R01}
}

@article{tabei06,
	author  = {Tabei, S. M. A. and Gingras, M. J. P. and Kao, Y.-J. and Stasiak, P. and Fortin, J.-Y.},
	title   = {Induced random fields in the {LiHo}$_x${Y}$_{1-x}${F}$_4$ quantum {I}sing magnet in a transverse magnetic field},
	journal = {Phys. Rev. Lett.},
	volume  = {97},
	pages   = {237203},
	year    = {2006},
	doi     = {10.1103/PhysRevLett.97.237203}
}

@article{schechter08,
	author  = {Schechter, Moshe},
	title   = {{LiHo}$_x${Y}$_{1-x}${F}$_4$ as a random-field {I}sing ferromagnet},
	journal = {Phys. Rev. B},
	volume  = {77},
	pages   = {020401},
	year    = {2008},
	doi     = {10.1103/PhysRevB.77.020401}
}

@article{silevitch19,
	author  = {Silevitch, D. M. and Tang, C. and Aeppli, G. and Rosenbaum, T. F.},
	title   = {Tuning high-{Q} nonlinear dynamics in a disordered quantum magnet},
	journal = {Nat. Commun.},
	volume  = {10},
	pages   = {4001},
	year    = {2019},
	doi     = {10.1038/s41467-019-11985-1}
}

@article{wen10,
	author  = {Wen, Bo and Subedi, P. and Bo, Lin and Yeshurun, Y. and Sarachik, M. P. and Kent, A. D. and Lampropoulos, C. and Christou, G.},
	title   = {Realization of random-field {I}sing ferromagnetism in a molecular magnet},
	journal = {Phys. Rev. B},
	volume  = {82},
	pages   = {014406},
	year    = {2010},
	doi     = {10.1103/PhysRevB.82.014406}
}

@article{aharony82,
	author  = {Aharony, A. and Gefen, Y. and Shapir, Y.},
	title   = {Dimensionality shift by three due to random fields in quantum spin systems},
	journal = {J. Phys. C: Solid State Phys.},
	volume  = {15},
	number  = {4},
	pages   = {673--679},
	year    = {1982},
	doi     = {10.1088/0022-3719/15/4/013}
}

@article{boyanovsky83,
	author  = {Boyanovsky, Daniel and Cardy, John L.},
	title   = {Dynamics of classical and quantum spin systems with random fields},
	journal = {Phys. Rev. B},
	volume  = {27},
	pages   = {5557--5564},
	year    = {1983},
	doi     = {10.1103/PhysRevB.27.5557}
}

@article{Chakrabot25,
  title = {Magnetic-field-tuned randomness in inhomogeneous altermagnets},
  author = {Chakraborty, Anzumaan R. and Schmalian, J\"org and Fernandes, Rafael M.},
  journal = {Phys. Rev. B},
  volume = {112},
  issue = {3},
  pages = {035146},
  numpages = {20},
  year = {2025},
  month = {Jul},
  publisher = {American Physical Society},
  doi = {10.1103/1vqq-9kzm},
  url = {https://link.aps.org/doi/10.1103/1vqq-9kzm}
}

@article{Gorkhov02,
  title = {Quantum collective creep: A quasiclassical Langevin equation approach},
  author = {Gorokhov, Denis A. and Fisher, Daniel S. and Blatter, Gianni},
  journal = {Phys. Rev. B},
  volume = {66},
  issue = {21},
  pages = {214203},
  numpages = {23},
  year = {2002},
  month = {Dec},
  publisher = {American Physical Society},
  doi = {10.1103/PhysRevB.66.214203},
  url = {https://link.aps.org/doi/10.1103/PhysRevB.66.214203}
}

@article{fisher86,
  title = {Scaling and critical slowing down in random-field Ising systems},
  author = {Fisher, Daniel S.},
  journal = {Phys. Rev. Lett.},
  volume = {56},
  issue = {5},
  pages = {416--419},
  numpages = {0},
  year = {1986},
  month = {Feb},
  publisher = {American Physical Society},
  doi = {10.1103/PhysRevLett.56.416},
  url = {https://link.aps.org/doi/10.1103/PhysRevLett.56.416}
}

@article{fisher86b,
  title = {Interface Fluctuations in Disordered Systems: $5-\epsilon$ Expansion and Failure of Dimensional Reduction},
  author = {Fisher, Daniel S.},
  journal = {Phys. Rev. Lett.},
  volume = {56},
  issue = {18},
  pages = {1964--1967},
  numpages = {0},
  year = {1986},
  month = {May},
  publisher = {American Physical Society},
  doi = {10.1103/PhysRevLett.56.1964},
  url = {https://link.aps.org/doi/10.1103/PhysRevLett.56.1964}
}

@article{tarjus04,
  title = {Nonperturbative Functional Renormalization Group for Random-Field Models: The Way Out of Dimensional Reduction},
  author = {Tarjus, Gilles and Tissier, Matthieu},
  journal = {Phys. Rev. Lett.},
  volume = {93},
  issue = {26},
  pages = {267008},
  numpages = {4},
  year = {2004},
  month = {Dec},
  publisher = {American Physical Society},
  doi = {10.1103/PhysRevLett.93.267008},
  url = {https://link.aps.org/doi/10.1103/PhysRevLett.93.267008}
}

@article{tarjus04_a,
  title = {Nonperturbative functional renormalization group for random field models and related disordered systems. I. Effective average action formalism},
  author = {Tarjus, Gilles and Tissier, Matthieu},
  journal = {Phys. Rev. B},
  volume = {78},
  issue = {2},
  pages = {024203},
  numpages = {19},
  year = {2008},
  month = {Jul},
  publisher = {American Physical Society},
  doi = {10.1103/PhysRevB.78.024203},
  url = {https://link.aps.org/doi/10.1103/PhysRevB.78.024203}
}

@article{tarjus04_b,
  title = {Nonperturbative functional renormalization group for random field models and related disordered systems. II. Results for the random field $O(N)$ model},
  author = {Tissier, Matthieu and Tarjus, Gilles},
  journal = {Phys. Rev. B},
  volume = {78},
  issue = {2},
  pages = {024204},
  numpages = {18},
  year = {2008},
  month = {Jul},
  publisher = {American Physical Society},
  doi = {10.1103/PhysRevB.78.024204},
  url = {https://link.aps.org/doi/10.1103/PhysRevB.78.024204}
}

@article{tissier11,
  title = {Supersymmetry and Its Spontaneous Breaking in the Random Field Ising Model},
  author = {Tissier, Matthieu and Tarjus, Gilles},
  journal = {Phys. Rev. Lett.},
  volume = {107},
  issue = {4},
  pages = {041601},
  numpages = {4},
  year = {2011},
  month = {Jul},
  publisher = {American Physical Society},
  doi = {10.1103/PhysRevLett.107.041601},
  url = {https://link.aps.org/doi/10.1103/PhysRevLett.107.041601}
}

@article{tissier12a,
  title = {Nonperturbative functional renormalization group for random field models and related disordered systems. III. Superfield formalism and ground-state dominance},
  author = {Tissier, Matthieu and Tarjus, Gilles},
  journal = {Phys. Rev. B},
  volume = {85},
  issue = {10},
  pages = {104202},
  numpages = {27},
  year = {2012},
  month = {Mar},
  publisher = {American Physical Society},
  doi = {10.1103/PhysRevB.85.104202},
  url = {https://link.aps.org/doi/10.1103/PhysRevB.85.104202}
}

@article{tissier12b,
  title = {Nonperturbative functional renormalization group for random field models and related disordered systems. IV. Supersymmetry and its spontaneous breaking},
  author = {Tissier, Matthieu and Tarjus, Gilles},
  journal = {Phys. Rev. B},
  volume = {85},
  issue = {10},
  pages = {104203},
  numpages = {17},
  year = {2012},
  month = {Mar},
  publisher = {American Physical Society},
  doi = {10.1103/PhysRevB.85.104203},
  url = {https://link.aps.org/doi/10.1103/PhysRevB.85.104203}
}

@article{tarjus13,
  title = {Avalanches and Dimensional Reduction Breakdown in the Critical Behavior of Disordered Systems},
  author = {Tarjus, Gilles and Baczyk, Maxime and Tissier, Matthieu},
  journal = {Phys. Rev. Lett.},
  volume = {110},
  issue = {13},
  pages = {135703},
  numpages = {5},
  year = {2013},
  month = {Mar},
  publisher = {American Physical Society},
  doi = {10.1103/PhysRevLett.110.135703},
  url = {https://link.aps.org/doi/10.1103/PhysRevLett.110.135703}
}

@book{MezardParisiVirasoro1987,
  author    = {M{\'e}zard, Marc and Parisi, Giorgio and Virasoro, Miguel Angel},
  title     = {Spin Glass Theory and Beyond},
  publisher = {World Scientific},
  address   = {Singapore},
  year      = {1987},
  doi       = {10.1142/0271}
}

@article{wiese22,
doi = {10.1088/1361-6633/ac4648},
url = {https://dx.doi.org/10.1088/1361-6633/ac4648},
year = {2022},
month = {aug},
publisher = {IOP Publishing},
volume = {85},
number = {8},
pages = {086502},
author = {Wiese, Kay J\"{o}rg},
title = {Theory and experiments for disordered elastic manifolds, depinning, avalanches, and sandpiles},
journal = {Rep. Prog. Phys.}
}

@article{balog_dynamics,
  author  = {I. Balog and G. Tarjus and M. Tissier},
  title   = {Criticality of the random field Ising model in and out of equilibrium: A nonperturbative functional renormalization group description},
  journal = {Phys. Rev. B},
  volume  = {97},
  pages   = {094204},
  year    = {2018},
  doi     = {10.1103/PhysRevB.97.094204}
}

@article{balog_activated,
  author  = {I. Balog and G. Tarjus},
  title   = {Activated dynamic scaling in the random-field Ising model: A nonperturbative functional renormalization group approach},
  journal = {Phys. Rev. B},
  volume  = {91},
  number  = {21},
  pages   = {214201},
  year    = {2015},
  doi     = {10.1103/PhysRevB.91.214201},
  url     = {https://doi.org/10.1103/PhysRevB.91.214201},
}

@article{dahmen96,
  title = {Hysteresis, avalanches, and disorder-induced critical scaling: A renormalization-group approach},
  author = {Dahmen, Karin and Sethna, James P.},
  journal = {Phys. Rev. B},
  volume = {53},
  issue = {22},
  pages = {14872--14905},
  numpages = {0},
  year = {1996},
  month = {Jun},
  publisher = {American Physical Society},
  doi = {10.1103/PhysRevB.53.14872},
  url = {https://link.aps.org/doi/10.1103/PhysRevB.53.14872}
}

@article{perkovic99,
  author  = {O. Perković and K. A. Dahmen and J. P. Sethna},
  title   = {Disorder-induced critical phenomena in hysteresis: Numerical scaling in three and higher dimensions},
  journal = {Phys. Rev. B},
  volume  = {59},
  pages   = {6106--6119},
  year    = {1999},
  doi     = {10.1103/PhysRevB.59.6106}
}

@article{sethna01,
  author  = {Sethna, James P. and Dahmen, Karin A. and Myers, Christopher R.},
  title   = {Crackling noise},
  journal = {Nature},
  year    = {2001},
  volume  = {410},
  number  = {6825},
  pages   = {242--250},
  doi     = {10.1038/35065675},
  url     = {https://doi.org/10.1038/35065675}
}

@article{wetterich93,
  author  = {Christof Wetterich},
  title   = {Exact evolution equation for the effective potential},
  journal = {Phys. Lett. B},
  volume  = {301},
  pages   = {90--94},
  year    = {1993},
  doi     = {10.1016/0370-2693(93)90726-X}
}

@article{berges02,
  author  = {J. Berges and N. Tetradis and C. Wetterich},
  title   = {Non-perturbative renormalization flow in quantum field theory and statistical physics},
  journal = {Phys. Rep.},
  volume  = {363},
  pages   = {223--386},
  year    = {2002},
  doi     = {10.1016/S0370-1573(01)00098-9}
}

@article{dupuis_review,
  author  = {N. Dupuis and L. Canet and A. Eichhorn and W. Metzner and J. M. Pawlowski and M. Tissier and N. Wschebor},
  title   = {The nonperturbative functional renormalization group and its applications},
  journal = {Phys. Rep.},
  volume  = {910},
  pages   = {1--114},
  year    = {2021},
  doi     = {10.1016/j.physrep.2021.01.001}
}

@article{imry-ma75,
  title = {Random-Field Instability of the Ordered State of Continuous Symmetry},
  author = {Imry, Yoseph and Ma, Shang-keng},
  journal = {Phys. Rev. Lett.},
  volume = {35},
  issue = {21},
  pages = {1399--1401},
  numpages = {0},
  year = {1975},
  month = {Nov},
  publisher = {American Physical Society},
  doi = {10.1103/PhysRevLett.35.1399},
  url = {https://link.aps.org/doi/10.1103/PhysRevLett.35.1399}
}

@article{aharony76,
  title = {Lowering of Dimensionality in Phase Transitions with Random Fields},
  author = {Aharony, Amnon and Imry, Yoseph and Ma, Shang-keng},
  journal = {Phys. Rev. Lett.},
  volume = {37},
  issue = {20},
  pages = {1364--1367},
  numpages = {0},
  year = {1976},
  month = {Nov},
  publisher = {American Physical Society},
  doi = {10.1103/PhysRevLett.37.1364},
  url = {https://link.aps.org/doi/10.1103/PhysRevLett.37.1364}
}

@article{grinstein76,
  title = {Ferromagnetic Phase Transitions in Random Fields: The Breakdown of Scaling Laws},
  author = {Grinstein, G.},
  journal = {Phys. Rev. Lett.},
  volume = {37},
  issue = {14},
  pages = {944--947},
  numpages = {0},
  year = {1976},
  month = {Oct},
  publisher = {American Physical Society},
  doi = {10.1103/PhysRevLett.37.944},
  url = {https://link.aps.org/doi/10.1103/PhysRevLett.37.944}
}

@article{parisi79,
  title = {Random Magnetic Fields, Supersymmetry, and Negative Dimensions},
  author = {Parisi, G. and Sourlas, N.},
  journal = {Phys. Rev. Lett.},
  volume = {43},
  issue = {11},
  pages = {744--745},
  numpages = {0},
  year = {1979},
  month = {Sep},
  publisher = {American Physical Society},
  doi = {10.1103/PhysRevLett.43.744},
  url = {https://link.aps.org/doi/10.1103/PhysRevLett.43.744}
}

@article{imbrie84,
  title = {Lower Critical Dimension of the Random-Field Ising Model},
  author = {Imbrie, John Z.},
  journal = {Phys. Rev. Lett.},
  volume = {53},
  issue = {18},
  pages = {1747--1750},
  numpages = {0},
  year = {1984},
  month = {Oct},
  publisher = {American Physical Society},
  doi = {10.1103/PhysRevLett.53.1747},
  url = {https://link.aps.org/doi/10.1103/PhysRevLett.53.1747}
}

@article{bricmont87,
  title = {Lower critical dimension for the random-field Ising model},
  author = {Bricmont, J. and Kupiainen, A.},
  journal = {Phys. Rev. Lett.},
  volume = {59},
  issue = {16},
  pages = {1829--1832},
  numpages = {0},
  year = {1987},
  month = {Oct},
  publisher = {American Physical Society},
  doi = {10.1103/PhysRevLett.59.1829},
  url = {https://link.aps.org/doi/10.1103/PhysRevLett.59.1829}
}

@article{villain84_b,
	author = {J. Villain},
	title = {Equilibrium critical properties of random field systems : new conjectures},
	DOI= "10.1051/jphys:0198500460110184300",
	url= "https://doi.org/10.1051/jphys:0198500460110184300",
	journal = {J. Phys. France},
	year = 1985,
	volume = 46,
	number = 11,
	pages = "1843-1852",
}

@article{aizenman89,
  author  = {Michael Aizenman and Jan Wehr},
  title   = {Rounding of First‐Order Phase Transitions in Systems with Quenched Disorder},
  journal = {Phys. Rev. Lett.},
  volume  = {62},
  number  = {21},
  pages   = {2503--2506},
  year    = {1989},
  doi     = {10.1103/PhysRevLett.62.2503},
}

@article{aizenman89_b,
  author  = {Michael Aizenman and Jan Wehr},
  title   = {Rounding Effects of Quenched Randomness on First‐Order Phase Transitions},
  journal = {Commun. Math. Phys.},
  volume  = {130},
  number  = {3},
  pages   = {489--528},
  year    = {1990},
  doi     = {10.1007/BF02096933},
}

@inbook{natterman98,
author = {T. Natterman},
title = {THEORY OF THE RANDOM FIELD ISING MODEL},
booktitle = {Spin Glasses and Random Fields},
chapter = {},
pages = {277-298},
doi = {10.1142/9789812819437_0009},
eprint ={cond-mat/9705295},
archivePrefix = {arXiv},
year={1998},
publisher={World Scientific},
address = {Singapore}
}

@article{birgenau98,
title = {Random fields and phase transitions in model magnetic systems},
journal = {J. Magn. Magn. Mater.},
volume = {177-181},
pages = {1-11},
year = {1998},
issn = {0304-8853},
doi = {https://doi.org/10.1016/S0304-8853(97)00998-0},
url = {https://www.sciencedirect.com/science/article/pii/S0304885397009980},
author = {R.J. Birgeneau}
}

@article{ogielski86,
  title = {Critical Behavior of the Three-Dimensional Dilute Ising Antiferromagnet in a Field},
  author = {Ogielski, Andrew T. and Huse, David A.},
  journal = {Phys. Rev. Lett.},
  volume = {56},
  issue = {12},
  pages = {1298--1301},
  numpages = {0},
  year = {1986},
  month = {Mar},
  publisher = {American Physical Society},
  doi = {10.1103/PhysRevLett.56.1298},
  url = {https://link.aps.org/doi/10.1103/PhysRevLett.56.1298}
}

@article{rieger93,
doi = {10.1088/0305-4470/26/20/014},
url = {https://doi.org/10.1088/0305-4470/26/20/014},
year = {1993},
month = {oct},
publisher = {},
volume = {26},
number = {20},
pages = {5279},
author = {H Rieger and A P Young},
title = {Critical exponents of the three-dimensional random field Ising model},
journal = {J. Phys. A: Math. Gen.}
}

@article{hartmann02,
  title = {Critical exponents of four-dimensional random-field Ising systems},
  author = {Hartmann, Alexander K.},
  journal = {Phys. Rev. B},
  volume = {65},
  issue = {17},
  pages = {174427},
  numpages = {8},
  year = {2002},
  month = {May},
  publisher = {American Physical Society},
  doi = {10.1103/PhysRevB.65.174427},
  url = {https://link.aps.org/doi/10.1103/PhysRevB.65.174427}
}

@article{middleton02,
  title = {Three-dimensional random-field Ising magnet: Interfaces, scaling, and the nature of states},
  author = {Middleton, A. Alan and Fisher, Daniel S.},
  journal = {Phys. Rev. B},
  volume = {65},
  issue = {13},
  pages = {134411},
  numpages = {31},
  year = {2002},
  month = {Mar},
  publisher = {American Physical Society},
  doi = {10.1103/PhysRevB.65.134411},
  url = {https://link.aps.org/doi/10.1103/PhysRevB.65.134411}
}

@article{litim01,
  author  = {Litim, Daniel F.},
  title   = {Optimized renormalization group flows},
  journal = {Phys. Rev. D},
  volume  = {64},
  pages   = {105007},
  year    = {2001},
  doi     = {10.1103/PhysRevD.64.105007}
}

@article{canet03,
  author  = {Canet, L{\'e}onie and Delamotte, Bertrand and Mouhanna, Dominique and Vidal, Julien},
  title   = {Optimization of the derivative expansion in the nonperturbative renormalization group},
  journal = {Phys. Rev. D},
  volume  = {67},
  pages   = {065004},
  year    = {2003},
  doi     = {10.1103/PhysRevD.67.065004},
  eprint  = {hep-th/0211055},
  archivePrefix = {arXiv}
}

@article{balog19,
  author  = {Balog, Ivan and Chat{\'e}, Hugues and Delamotte, Bertrand and Marohni{\'c}, Maroje and Wschebor, Nicol{\'a}s},
  title   = {Convergence of Non-Perturbative Approximations to the Renormalization Group},
  journal = {Phys. Rev. Lett.},
  volume  = {123},
  pages   = {240604},
  year    = {2019},
  doi     = {10.1103/PhysRevLett.123.240604},
  eprint  = {1907.01829},
  archivePrefix = {arXiv}
}

@article{depolsi22,
  author  = {De Polsi, Gonzalo and Wschebor, Nicol{\'a}s},
  title   = {Regulator dependence in the functional renormalization group: A quantitative explanation},
  journal = {Phys. Rev. E},
  volume  = {106},
  pages   = {024111},
  year    = {2022},
  doi     = {10.1103/PhysRevE.106.024111}
}

@Article{tarjus25,
	title={{On the breakdown of dimensional reduction and supersymmetry in random-field models}},
	author={Gilles Tarjus and Matthieu Tissier and Ivan Balog},
	journal={SciPost Phys.},
	volume={19},
	pages={001},
	year={2025},
	publisher={SciPost},
	doi={10.21468/SciPostPhys.19.1.001},
	url={https://scipost.org/10.21468/SciPostPhys.19.1.001},
}

@article{andreanov14,
  title = {Localization of spin waves in disordered quantum rotors},
  author = {Andreanov, Alexei and Fedorenko, Andrei A.},
  journal = {Phys. Rev. B},
  volume = {90},
  issue = {1},
  pages = {014205},
  numpages = {8},
  year = {2014},
  month = {Jul},
  publisher = {American Physical Society},
  doi = {10.1103/PhysRevB.90.014205},
  url = {https://link.aps.org/doi/10.1103/PhysRevB.90.014205}
}

@article{daviet20,
  title = {Mott-Glass Phase of a One-Dimensional Quantum Fluid with Long-Range Interactions},
  author = {Daviet, Romain and Dupuis, Nicolas},
  journal = {Phys. Rev. Lett.},
  volume = {125},
  issue = {23},
  pages = {235301},
  numpages = {6},
  year = {2020},
  month = {Dec},
  publisher = {American Physical Society},
  doi = {10.1103/PhysRevLett.125.235301},
  url = {https://link.aps.org/doi/10.1103/PhysRevLett.125.235301}
}

@article{dupuis24,
  title = {Mott-glass phase induced by long-range correlated disorder in a one-dimensional Bose gas},
  author = {Dupuis, Nicolas and Fedorenko, Andrei A.},
  journal = {Phys. Rev. A},
  volume = {110},
  issue = {4},
  pages = {L041304},
  numpages = {6},
  year = {2024},
  month = {Oct},
  publisher = {American Physical Society},
  doi = {10.1103/PhysRevA.110.L041304},
  url = {https://link.aps.org/doi/10.1103/PhysRevA.110.L041304}
}

@article{grison26,
  title = {From Bose glass to many-body localization in a one-dimensional disordered Bose gas},
  author = {Grison, Vincent and Dupuis, Nicolas},
  journal = {Phys. Rev. B},
  volume = {113},
  issue = {17},
  pages = {174208},
  numpages = {18},
  year = {2026},
  month = {May},
  publisher = {American Physical Society},
  doi = {10.1103/6cby-152d},
  url = {https://link.aps.org/doi/10.1103/6cby-152d}
}

@article{halperin72,
  title = {Calculation of Dynamic Critical Properties Using Wilson's Expansion Methods},
  author = {Halperin, B. I. and Hohenberg, P. C. and Ma, Shang-keng},
  journal = {Phys. Rev. Lett.},
  volume = {29},
  issue = {23},
  pages = {1548--1551},
  numpages = {0},
  year = {1972},
  month = {Dec},
  publisher = {American Physical Society},
  doi = {10.1103/PhysRevLett.29.1548},
  url = {https://link.aps.org/doi/10.1103/PhysRevLett.29.1548}
}

@article{hohenberg77,
  title = {Theory of dynamic critical phenomena},
  author = {Hohenberg, P. C. and Halperin, B. I.},
  journal = {Rev. Mod. Phys.},
  volume = {49},
  issue = {3},
  pages = {435--479},
  numpages = {0},
  year = {1977},
  month = {Jul},
  publisher = {American Physical Society},
  doi = {10.1103/RevModPhys.49.435},
  url = {https://link.aps.org/doi/10.1103/RevModPhys.49.435}
}

@article{schwartz85,
  title = {Exact Inequality for Random Systems: Application to Random Fields},
  author = {Schwartz, Moshe and Soffer, A.},
  journal = {Phys. Rev. Lett.},
  volume = {55},
  issue = {22},
  pages = {2499--2501},
  numpages = {0},
  year = {1985},
  month = {Nov},
  publisher = {American Physical Society},
  doi = {10.1103/PhysRevLett.55.2499},
  url = {https://link.aps.org/doi/10.1103/PhysRevLett.55.2499}
}

@article{tarjus13a,
doi = {10.1209/0295-5075/103/61001},
url = {https://doi.org/10.1209/0295-5075/103/61001},
year = {2013},
month = {oct},
publisher = {EDP Sciences, IOP Publishing and Società Italiana di Fisica},
volume = {103},
number = {6},
pages = {61001},
author = {Tarjus, Gilles and Balog, Ivan and Tissier, Matthieu},
title = {Critical scaling in random-field systems: 2 or 3 independent exponents?},
journal = {Europhys. Lett.}
}

@article{chauve01,
  author  = {Chauve, Pascal and Le Doussal, Pierre and Wiese, Kay J{\"o}rg},
  title   = {Renormalization of Pinned Elastic Systems: How Does It Work beyond One Loop?},
  journal = {Phys. Rev. Lett.},
  volume  = {86},
  pages   = {1785--1788},
  year    = {2001},
  doi     = {10.1103/PhysRevLett.86.1785},
  eprint  = {cond-mat/0006056},
  archivePrefix = {arXiv}
}

@article{ledoussal02_dep,
  author  = {Le Doussal, Pierre and Wiese, Kay J{\"o}rg and Chauve, Pascal},
  title   = {Two-loop functional renormalization group theory of the depinning transition},
  journal = {Phys. Rev. B},
  volume  = {66},
  pages   = {174201},
  year    = {2002},
  doi     = {10.1103/PhysRevB.66.174201},
  eprint  = {cond-mat/0205108},
  archivePrefix = {arXiv}
}

@article{ledoussal08,
  author  = {Le Doussal, Pierre and Mueller, Markus and Wiese, Kay J{\"o}rg},
  title   = {Cusps and shocks in the renormalized potential of glassy random manifolds: How functional renormalization group and replica symmetry breaking fit together},
  journal = {Phys. Rev. B},
  volume  = {77},
  pages   = {064203},
  year    = {2008},
  doi     = {10.1103/PhysRevB.77.064203},
  eprint  = {0711.3929},
  archivePrefix = {arXiv}
}

@article{fisher88,
  author  = {Fisher, Daniel S. and Huse, David A.},
  title   = {Nonequilibrium dynamics of spin glasses},
  journal = {Phys. Rev. B},
  volume  = {38},
  pages   = {373--385},
  year    = {1988},
  doi     = {10.1103/PhysRevB.38.373}
}

@article{braymoore87,
  author  = {Bray, A. J. and Moore, M. A.},
  title   = {Chaotic nature of the spin-glass phase},
  journal = {Phys. Rev. Lett.},
  volume  = {58},
  pages   = {57--60},
  year    = {1987},
  doi     = {10.1103/PhysRevLett.58.57}
}

@article{lubchenko07,
  author  = {Lubchenko, Vassiliy and Wolynes, Peter G.},
  title   = {Theory of structural glasses and supercooled liquids},
  journal = {Annu. Rev. Phys. Chem.},
  volume  = {58},
  pages   = {235--266},
  year    = {2007},
  doi     = {10.1146/annurev.physchem.58.032806.104653}
}

@article{bouchaud98,
  author  = {Bouchaud, Jean-Philippe and Cugliandolo, Leticia F. and Kurchan, Jorge and M{\'e}zard, Marc},
  title   = {Out of equilibrium dynamics in spin-glasses and other glassy systems},
  journal = {Spin Glasses and Random Fields},
  pages   = {161--223},
  year    = {1998},
  publisher = {World Scientific},
  eprint  = {cond-mat/9702070},
  archivePrefix = {arXiv}
}

@article{herz76,
  title = {Quantum critical phenomena},
  author = {Hertz, John A.},
  journal = {Phys. Rev. B},
  volume = {14},
  issue = {3},
  pages = {1165--1184},
  numpages = {0},
  year = {1976},
  month = {Aug},
  publisher = {American Physical Society},
  doi = {10.1103/PhysRevB.14.1165},
  url = {https://link.aps.org/doi/10.1103/PhysRevB.14.1165}
}

@article{anfuso09,
  author  = {Anfuso, F. and Rosch, A.},
  title   = {Random field effects in field-driven quantum critical points},
  journal = {Eur. Phys. J. B},
  volume  = {69},
  number  = {4},
  pages   = {465--471},
  year    = {2009},
  doi     = {10.1140/epjb/e2009-00191-6},
  url     = {https://doi.org/10.1140/epjb/e2009-00191-6}
}

@article{depolsi20,
  title = {Precision calculation of critical exponents in the O(N) universality classes with the nonperturbative renormalization group},
  author = {De Polsi, Gonzalo and Balog, Ivan and Tissier, Matthieu and Wschebor, Nicol\'as},
  journal = {Phys. Rev. E},
  volume = {101},
  issue = {4},
  pages = {042113},
  numpages = {22},
  year = {2020},
  month = {Apr},
  publisher = {American Physical Society},
  doi = {10.1103/PhysRevE.101.042113},
  url = {https://link.aps.org/doi/10.1103/PhysRevE.101.042113}
}

@article{chauve00,
  author = {Chauve, P. and Giamarchi, T. and Le Doussal, P.},
  title = {Creep and depinning in disordered media},
  journal = {Phys. Rev. B},
  volume = {62},
  pages = {6241--6267},
  year = {2000}
}

@article{chang25,
  author        = {Chang, Cyuan-Han and Dommes, Vasiliy and Erramilli, Rajeev S. and Homrich, Alexandre and Kravchuk, Petr and Liu, Aike and Mitchell, Matthew S. and Poland, David and Simmons-Duffin, David},
  title         = {Bootstrapping the 3d Ising stress tensor},
  journal       = {JHEP},
  volume        = {2025},
  number        = {3},
  pages         = {136},
  year          = {2025},
  doi           = {10.1007/JHEP03(2025)136}
}

@article{fytas18,
  author        = {Fytas, Nikolaos G. and Mart{\'i}n-Mayor, V{\'i}ctor and Picco, Marco and Sourlas, Nicolas},
  title         = {Review of recent developments in the random-field Ising model},
  journal       = {J. Stat. Phys.},
  volume        = {172},
  pages         = {665--672},
  year          = {2018},
  doi           = {10.1007/s10955-018-1955-7},
  eprint        = {1711.09597},
  archivePrefix = {arXiv},
  primaryClass  = {cond-mat.dis-nn}
}

@misc{balog19a,
      title={Comment on "Evidence for Supersymmetry in the Random-Field Ising Model at D=5''}, 
      author={Ivan Balog and Gilles Tarjus and Matthieu Tissier},
      year={2019},
      eprint={1906.10058},
      archivePrefix={arXiv},
      primaryClass={cond-mat.dis-nn},
      url={https://arxiv.org/abs/1906.10058}, 
}

@article{suzuki76,
  author  = {Suzuki, Masuo},
  title   = {Relationship between d-Dimensional Quantal Spin Systems and (d+1)-Dimensional Ising Systems: Equivalence, Critical Exponents and Systematic Approximants of the Partition Function and Spin Correlations},
  journal = {Prog. Theor. Phys.},
  volume  = {56},
  number  = {5},
  pages   = {1454--1469},
  year    = {1976},
  doi     = {10.1143/PTP.56.1454}
}

@article{homenda24,
  title = {Generalized Hertz action and quantum criticality of two-dimensional Fermi systems},
  author = {Homenda, Mateusz and Jakubczyk, Pawel and Yamase, Hiroyuki},
  journal = {Phys. Rev. B},
  volume = {110},
  issue = {12},
  pages = {L121102},
  numpages = {6},
  year = {2024},
  month = {Sep},
  publisher = {American Physical Society},
  doi = {10.1103/PhysRevB.110.L121102},
  url = {https://link.aps.org/doi/10.1103/PhysRevB.110.L121102}
}

@misc{homenda25,
      title={Quantum criticality and non-Fermi liquids: the nonperturbative renormalization group perspective}, 
      author={Mateusz Homenda and Pawel Jakubczyk and Hiroyuki Yamase},
      year={2025},
      eprint={2505.10140},
      archivePrefix={arXiv},
      primaryClass={cond-mat.str-el},
      url={https://arxiv.org/abs/2505.10140}, 
}

@article{nash91,
  title = {Experimental verification of activated critical dynamics in the d=3 random-field Ising model},
  author = {Nash, A. E. and King, A. R. and Jaccarino, V.},
  journal = {Phys. Rev. B},
  volume = {43},
  issue = {1},
  pages = {1272(R)--1275(R)},
  numpages = {0},
  year = {1991},
  month = {Jan},
  publisher = {American Physical Society},
  doi = {10.1103/PhysRevB.43.1272},
  url = {https://link.aps.org/doi/10.1103/PhysRevB.43.1272}
}

@article{lederman92,
  author  = {Lederman, M. and Selinger, J. V. and Bruinsma, R. and Hammann, J. and Orbach, R.},
  title   = {Low-temperature dynamics of a diluted Ising antiferromagnet},
  journal = {Phys. Rev. Lett.},
  volume  = {68},
  pages   = {2086--2089},
  year    = {1992},
  doi     = {10.1103/PhysRevLett.68.2086}
}

@incollection{belanger97,
  author        = {Belanger, D. P.},
  title         = {Experiments on the random field Ising model},
  booktitle     = {Spin Glasses and Random Fields},
  editor        = {Young, A. P.},
  publisher     = {World Scientific},
  year          = {1998},
  eprint        = {cond-mat/9706042},
  archivePrefix = {arXiv}
}

\end{document}